\documentclass[11pt,a4paper]{article}
\pdfoutput=1
\usepackage{a4wide}
\usepackage{amsmath}
\usepackage{amssymb}
\usepackage{amsfonts}
\usepackage{cite}
\usepackage[table]{xcolor}
\usepackage{adjustbox}
\usepackage{multirow}
\usepackage{pdflscape}
\usepackage{gensymb}
\usepackage{graphicx}
\usepackage{diagbox}
\usepackage{pifont}
\usepackage{bigstrut}
\usepackage[small,bf]{caption}
\usepackage{enumerate}
\usepackage{tabulary}
\usepackage{float}
\usepackage{arydshln}
\usepackage{enumitem}
\usepackage{mathtools}
\usepackage[utf8]{inputenc}
\usepackage[T1]{fontenc}
\usepackage{footnote}
\usepackage{footmisc}
\usepackage{hyperref}
\usepackage{ctable}
\def\1{\mathbf{1}}
\def\3{\mathbf{3}}
\def\2{\mathbf{2}}

\DeclarePairedDelimiter{\abs}{\lvert}{\rvert}
\numberwithin{equation}{section}
\makeatletter
\g@addto@macro\bfseries{\boldmath}
\makeatother

\newcounter{savefootnote}
\newcounter{symfootnote}
\newcommand{\symfootnote}[1]{%
   \setcounter{savefootnote}{\value{footnote}}%
   \setcounter{footnote}{\value{symfootnote}}%
   \ifnum\value{footnote}>8\setcounter{footnote}{0}\fi%
   \let\oldthefootnote=\thefootnote%
   \renewcommand{\thefootnote}{\fnsymbol{footnote}}%
   \footnote{#1}%
   \let\thefootnote=\oldthefootnote%
   \setcounter{symfootnote}{\value{footnote}}%
   \setcounter{footnote}{\value{savefootnote}}%
}

\begin{document}
%=============

\begin{titlepage}

\vspace*{-15mm}

\begin{center}

{\bf\LARGE{Implications of the Dark-LMA solution for neutrino mass matrices}}\\[10mm]

Monojit Ghosh,$^{a,b,c}$\footnote{E-mail: \texttt{manojit@kth.se}} 
Srubabati Goswami,$^{d,}$\footnote{E-mail: \texttt{sruba@prl.res.in}} 
and
Ananya Mukherjee$^{\,d,}$\footnote{E-mail: \texttt{ananya@prl.res.in}} \\
\vspace{8mm}
$^{a}$\,{\it Department of Physics, School of Engineering Sciences,\\ KTH Royal Institute of Technology, AlbaNova University Center,\\ Roslagstullsbacken 21, SE--106 91 Stockholm, Sweden} \\
\vspace{2mm}
$^{b}$\,{\it The Oskar Klein Centre for Cosmoparticle Physics, AlbaNova University Center, Roslagstullsbacken 21, SE--106 91 Stockholm, Sweden} \\
\vspace{2mm}
$^{c}$\,{\it Center of Excellence for Advanced Materials and Sensing Devices, Ruder Bo\v{s}kovi\'c Institute, 10000 Zagreb, Croatia}\\
\vspace{2mm}
$^{d}$\,{\it Physical Research Laboratory, Ahmedabad- 380009, India} \\

\end{center}
\vspace{8mm}

\begin{abstract}
\noindent 
In this work we have re-investigated two different kinds of texture zero ansatz of the low energy neutrino mass matrix  in view of the Dark-Large-Mixing-Angle (DLMA) solution of the solar neutrino problem which can arise in the presence of non-standard interactions.
In particular we revisit the cases of (i) one zero mass matrices when the lowest neutrino mass is zero and (ii) one zero texture with a vanishing minor. In our study we find that for most of the cases, the texture zero conditions which are allowed for the LMA solution, are also allowed for the DLMA solution. However, we found two textures belonging to the case of one zero texture with a vanishing minor where LMA solution does not give a viable solution whereas DLMA solution does. We analyze all the possible texture zero cases belonging to these two kinds of texture zero structures in detail and present correlations between different parameters. We also present the predictions for the effective neutrino mass governing neutrino-less double beta decay for the allowed textures.
\end{abstract}
\end{titlepage}

\setcounter{footnote}{0}

\section{Introduction}
With neutrino physics being in the precision era, the determination of
neutrino observables and the associated theoretical studies have become one of the prime objectives in astroparticle and high energy physics. The unparalleled effort from various oscillation experiments has greatly helped in the precise determination of the neutrino parameters. As a result we have the information on the magnitude of two non zero neutrino mass squared splittings ($\Delta m_{ij}^2$) and three mixing angles ($\theta_{ij}$) with a very good accuracy. However, a precise measurement of the value of $\delta_{\text{CP}}$, the true octant of the atmospheric mixing angle $\theta_{23}$ and determination of the neutrino mass ordering still need more efforts. The measurement of all the above mentioned parameters are based only on the leading effects which describe the interaction among the light neutrinos and the matter fields via standard interaction only. But, these measurements can go through certain obscurities if one includes the non standard interaction (NSI) of light neutrinos with the matter field \cite{PhysRevD.17.2369,Guzzo:1991hi} along with the standard interactions. Experimental efforts are underway to investigate the sub-leading non-oscillation effects such as the NSI. 
%NSI$DLMA
On account of the presence of NSI, solar neutrino data admit a new solution corresponding to $\theta_{12} > 45^0$, which is known as dark large mixing angle (DLMA) solution \cite{Miranda:2004nb,Escrihuela:2010zz,Farzan:2017xzy}. The standard large mixing angle (LMA) solution is nearly degenerate with this DLMA solution for $\Delta m_{21}^2 \sim 7.5 \times 10^{-5}eV^2$ and $\sin^2 \theta_{12} \sim 0.7$.  Very recently neutrino-nucleus scattering data from COHERENT experiment has been found to constrain the DLMA parameter space severely~ \cite{Esteban:2018ppq}.
However these bounds are model dependent and  it depends on the mass of the light mediator \cite{Denton:2018xmq}.  In this regard it is to mention here that COHERENT data exclude the DLMA solution at 95 $\%$ C.L. if the light mediator mass is greater than 48 MeV only. However, the global analysis including oscillation and COHERENT data says that the DLMA solution can still be allowed at $3\sigma$, when the NSI parameters have a smaller range of values and with light mediators of mass $\geq$10MeV. There have been various theoretical studies in regard to this alternate solution to the solar mixing angle. It is shown that this DLMA solution is the manifestation of a generalized hierarchy degeneracy \cite{Farzan:2017xzy,Bakhti:2014pva,Coloma:2016gei,Choubey:2019osj}. In the study of neutrinoless double beta decay, impact of the DLMA solution on the effective neutrino mass has been highlighted in \cite{PhysRevD.99.095038,Deepthi:2019ljo,Ge:2019ldu}. 

In the last decade a plenty of theoretical studies have been carried out to understand the neutrino mixing pattern and hence the underlying symmetry of the low energy neutrino mass matrix. Texture zero is example of one of such studies. Textures imply certain relationships among leptonic mixing parameters leading to the possibility of having vanishing elements/minors in the neutrino mass matrix. Study of texture zero in the standard three flavour scenario has been carried out in great detail in light of the LMA solution. We refer to \cite{Ludl:2014axa} for a comprehensive review. However the studies of texture zero in the context of the DLMA solution has not received much attention. With the emergence of the DLMA solution, it is worthwhile to revisit the texture zero in light of these new solutions as the results obtained for the LMA solution may change in the presence of the DLMA solution. 

In this work we re-examine texture zero scenarios of the neutrino mass matrix in light of the DLMA solution, where (i) one or more elements of the low energy neutrino mass matrix can be zero and (ii) one of the elements of neutrino mass matrix and its minor is zero simultaneously. Recently ref. \cite{Borgohain:2020now} has carried out the analysis of (i) when the lowest neutrino mass is non zero. In this work, using the most recent data, the authors confirm the fact that all the possible six one-zero textures are allowed \cite{Merle:2006du,PhysRevD.85.113011,Dev:2006if,Deepthi:2011sk} and among the possible 15 two-zero textures only seven are allowed \cite{Frampton:2002yf,Xing:2002ta,Guo:2002ei,Xing:2003ic,Grimus:2004az,Xing:2004ik,Kaneko:2005yz,Dev:2006qe,Kumar:2011vf,Fritzsch:2011qv,Meloni:2012sx,Meloni:2014yea,Zhou:2015qua,Cebola:2015dwa,Nishi:2016wki,Singh:2016qcf} for the LMA solution. More than two zero textures are not allowed in standard three flavour framework. In addition they find that all the one-zero textures and all the two zero textures which are allowed with the LMA solution are also allowed with the DLMA solution except the ones with vanishing $m_{ee}$. The cases with $m_{ee}=0$ which are allowed with the LMA solution, are disfavoured with the DLMA solution. In our present work we consider the same one-zero textures but when the lowest neutrino mass is zero\footnote{We have checked that two zero textures are not allowed when lowest neutrino mass is zero.}.The theoretical motivation for considering the lowest neutrino mass to be zero goes as follows.

As we know in the type-I seesaw model, the effective
neutrino mass matrix is described by $M_\nu \approx  M_D M_R^{-1}M_D^T$, which requires at least two right handed neutrinos (RHN) to be consistent with the non-zero neutrino mass squared differences (solar and atmospheric). The addition of only two RHNs to the  Standard Model brings an economical extension~\cite{Ma:1998zg,Ibarra:2003up,Ibarra:2005qi}. It is true that, presence of only two RHNs in the type-I seesaw model predicts one of the neutrino mass eigenvalue to be zero. Therefore, driven by this we have reconsidered a vanishing lowest neutrino mass.
The study of one zero texture with a vanishing minor in light of LMA solution is carried out in \cite{Dev:2010if,Liao:2013saa}. These kind of textures can be realized via seesaw mechanism when some of the elements of $M_D$ and $M_R$ are zero.

We organize this paper in the following manner. In Section~\ref{lowenergy} we describe the low energy neutrino mass matrix and the relevant parameters which delineate neutrino oscillation. Section~\ref{OZT} is kept for the new findings and analysis on one zero texture study with the present backgrounds for the solar mixing angle. In Section~\ref{TCS} we present the results and analysis on the neutrino mass pattern followed by the simultaneous existence of one zero texture and a vanishing minor. In Section \ref{ndbd}, we present the predictions of the effective neutrino mass $m_{\beta\beta}$ for the allowed textures. Finally we draw our conclusion in Section~\ref{conclusion}.
 %%%%%%%%%%%%%%      
\section{Low energy neutrino mass matrix}  \label{lowenergy}
The well known lepton mixing matrix $U_{\text{PMNS}}$ represents the mixing between the neutrino flavour eigenstates and their mass eigenstates. In a three flavored paradigm this matrix is parameterised in terms of three mixing angles and three CP phases as,
\begin{equation}
V_{\text{PMNS}} =  U_{\text{PMNS}} \,U_{\text{Maj}}
\end{equation} 
where
\begin{equation}
U_{\text{PMNS}}=\left(\begin{array}{ccc}
c_{12}c_{13}& s_{12}c_{13}& s_{13}e^{-i\delta_{\rm CP}}\\
-s_{12}c_{23}-c_{12}s_{23}s_{13}e^{i\delta}& c_{12}c_{23}-s_{12}s_{23}s_{13}e^{i\delta} & s_{23}c_{13} \\
s_{12}s_{23}-c_{12}c_{23}s_{13}e^{i\delta} & -c_{12}s_{23}-s_{12}c_{23}s_{13}e^{i\delta}& c_{23}c_{13}
\end{array}\right),
\label{PMNS}
\end{equation}
with $c_{ij} = \cos{\theta_{ij}}, \; s_{ij} = \sin{\theta_{ij}}$ and $\delta_{\rm CP}$ as the Dirac CP phase. The diagonal
matrix, $U_{\text{Maj}}=\text{diag}(1, e^{-i\alpha}, e^{i(-\beta+\delta)})$, contains the 
Majorana CP phases $\alpha, \beta$ which can only be probed in the neutrinoless double beta decay experiments and not in neutrino oscillation experiments.

With a diagonal charged lepton mass matrix, the low energy neutrino mass matrix in the flavor basis, can be expressed with the help of the lepton mixing matrix Eq. \ref{PMNS} as,
\begin{equation}
M_\nu = V_{\text{PMNS}}M_{\nu}^{\text{diag}}V_{\text{PMNS}}^T,
\end{equation}
where, $M_{\nu}^{\text{diag}}$ carries the neutrino mass eigenvalues $m_1,~ m_2, ~ m_3$. Depending on the choice of the neutrino mass hierarchy we can express the mass eigenvalues in terms of the solar ($\Delta m_{\text{sol}}^2$) and atmospheric ($\Delta m_{\text{atm}}^2$) mass splittings as $m_1, m_2=\sqrt{m_1^2 + \Delta m_{\text{sol}}^2}, m_3=\sqrt{m_1^2 + \Delta m_{\text{atm}}^2}$ for normal hierarchy (NH) of the neutrino masses whereas for inverted hierarchy (IH) of the neutrino masses $m_1 = \sqrt{m_3^2 + \Delta m_{\text{atm}}^2}, ~m_2=\sqrt{m_3^2 + \Delta m_{\text{sol}}^2+ \Delta m_{\text{atm}}^2}, m_3$.
For NH, one can approximate the mass eigenvalues as $\abs{m_3}\simeq \Delta m_{\text{atm}}^2 \gg \abs{m_2}\simeq \Delta m_{\text{sol}}^2 \gg \abs{m_1}$ and $\abs{m_2}\simeq \abs{m_1}\simeq \Delta m_{\text{atm}}^2 \gg \abs{m_3}$ for IH.

As mentioned earlier, we re-investigate here the one zero texture scheme with vanishing lowest neutrino mass and  one zero texture with vanishing minor considering a non-zero lowest neutrino mass with the latest global fit for neutrino oscillation parameters along with the second choice of the solar mixing angle which is also known as Dark-LMA solution. The recent $3 \sigma$ global fit of the three flavor oscillation parameters \cite{Esteban:2018azc} can be found from the Table \ref{tabdata}. The Dark-LMA values allow us to write the other solution for the solar angle as $\sin^2\theta_{12} (\text{DLMA}) = 0.650 - 0.725$ \cite{Farzan:2017xzy}. It is to be mentioned here that for numerical analysis 
we have chosen the entire $3\sigma$ ranges of the oscillation parameters and $\delta_{\text{CP}}$ varied from $0$ to $360^\circ$. Taking these choices for oscillation parameters as inputs and imposing the texture zero conditions on the mass matrix elements we classify the allowed texture classes and study the correlations for these cases.
 It is to be noted that the global analysis including NSI predicts the ranges of $\sin^2\theta_{12} = 0.214-0.356$ (LMA), $\sin^2\theta_{12} = 0.648 - 0.745$ (DLMA), $\Delta m^2_{21} = (6.73 - 8.14) \times 10^{-5}$ (LMA) and $\Delta m^2_{21} = (6.82 - 8.02) \times 10^{-5}$ (DLMA) while the other oscillation parameters remain unaffected. Since the change is very marginal \cite{Esteban:2018ppq}, we have used the ranges in Table \ref{tabdata} in our analysis.
 \begin{table}[h!]
%\hspace{-1.186 in}
%\renewcommand*{\arraystretch}{1.5}
\begin{center}
\begin{tabular}{|c|c|c|}
\hline
Parameters & Normal ordering & Inverted ordering  \\
\hline
$\sin^2\theta_{23}$ & 0.433- 0.609 & 0.436 - 0.610\\
\hline
$\sin^2\theta_{12}$ & 0.275- 0.350 & 0.275- 0.350 \\
\hline
 $\sin^2\theta_{13}$ & 0.02044- 0.02435 & 0.02064 - 0.02457 \\
\hline
$\Delta m^2_{21}$ & $(6.79 -8.01 )\times 10^{-5}$ eV$^2$ & $(6.79-8.01 ) \times 10^{-5}$ eV$^2$\\
\hline
$\Delta m^2_{31}$ & $(2.436-2.618) \times 10^{-3}$ eV$^2$ & $-(2.601 - 2.419) \times 10^{-3}$ eV$^2$\\
\hline
$\delta_{CP}/ ^{0}$ &144-357 & 205-348 \\
\hline
\end{tabular}
\caption{Latest $3\sigma$ bounds on the oscillation parameters from Ref. \cite{Esteban:2018azc}.}
\label{tabdata}
\end{center}
\end{table}
 
\section{One-zero textures}\label{OZT}

In this section we present the viable cases in the context of one-zero texture neutrino mass matrices when the lowest neutrino mass is zero. 
\begin{figure*}[th!]
\begin{center}
\includegraphics[scale=0.41]{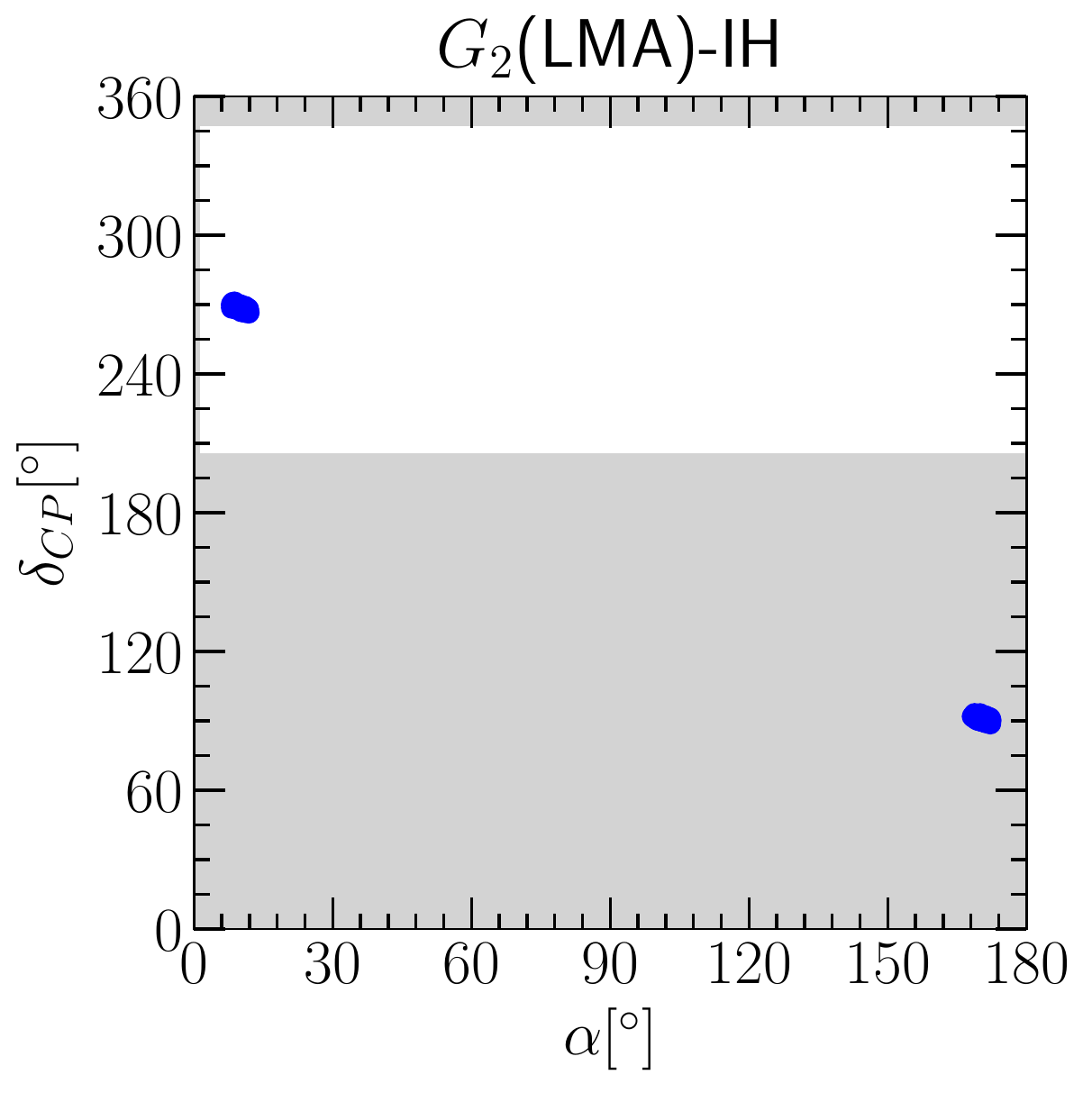}
\hspace{0.8 in}   
\includegraphics[scale=0.41]{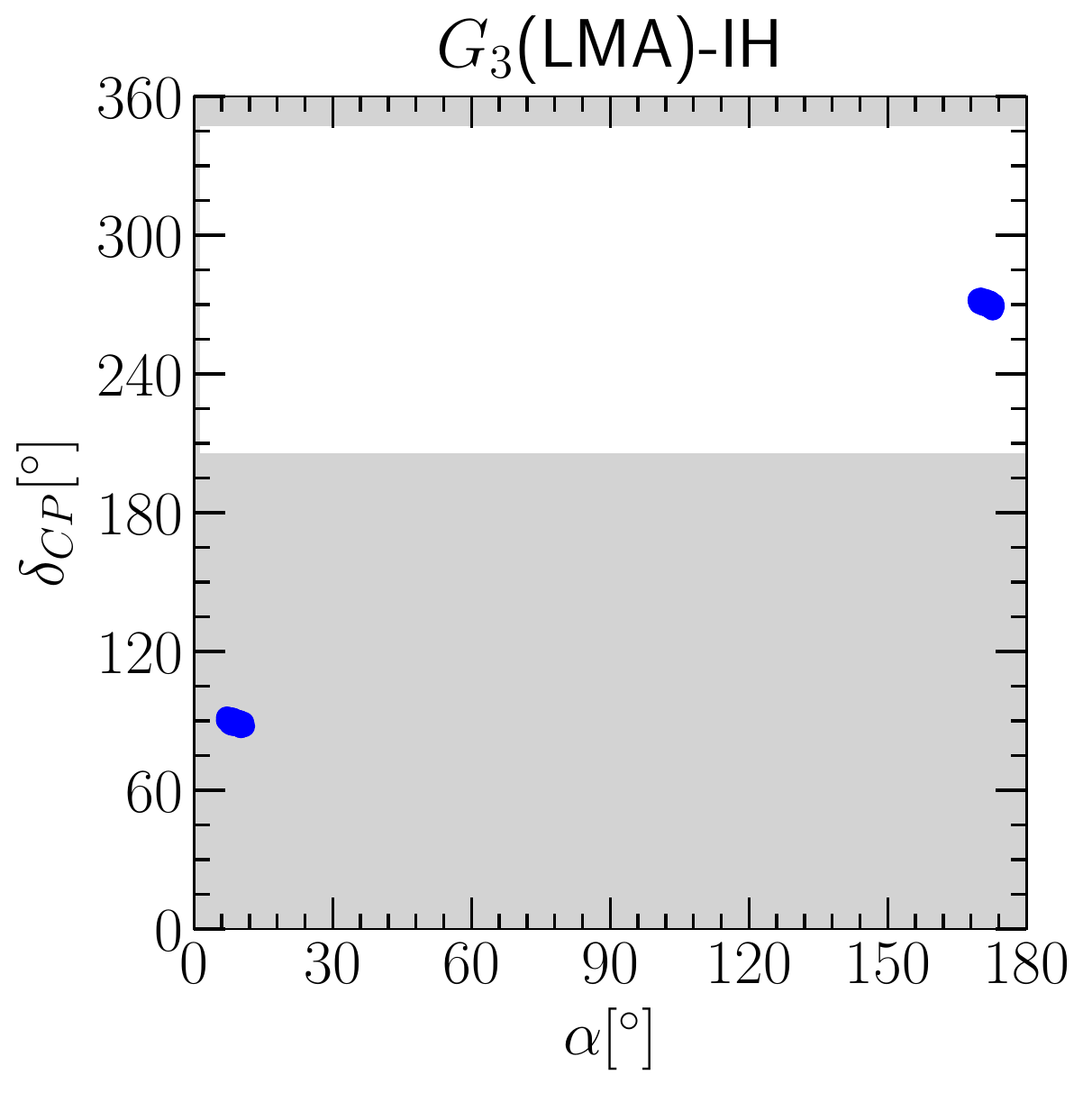} \\     
\includegraphics[scale=0.4]{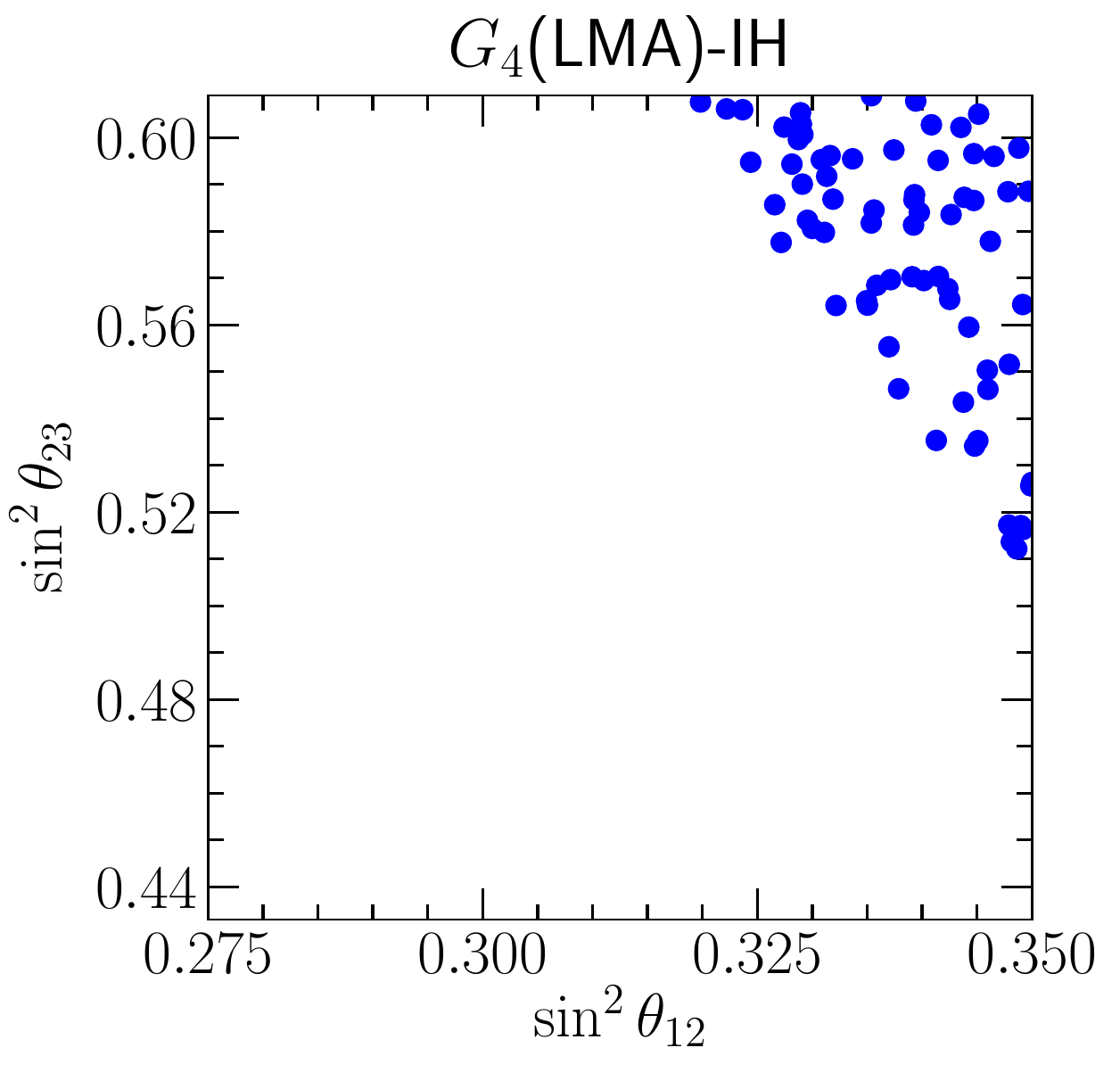}
\hspace{0.8 in}
\includegraphics[scale=0.4]{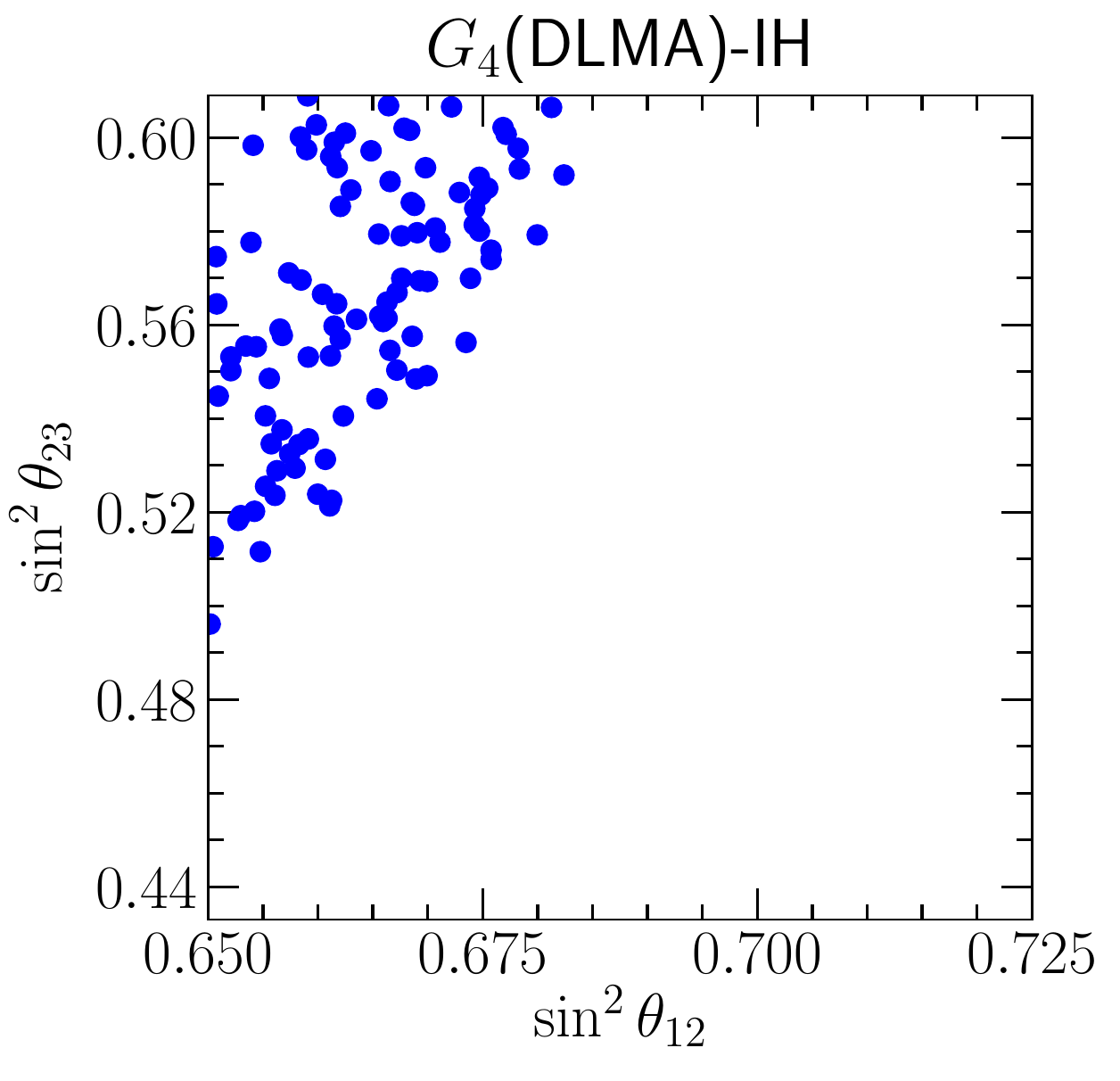}   \\  
\includegraphics[scale=0.41]{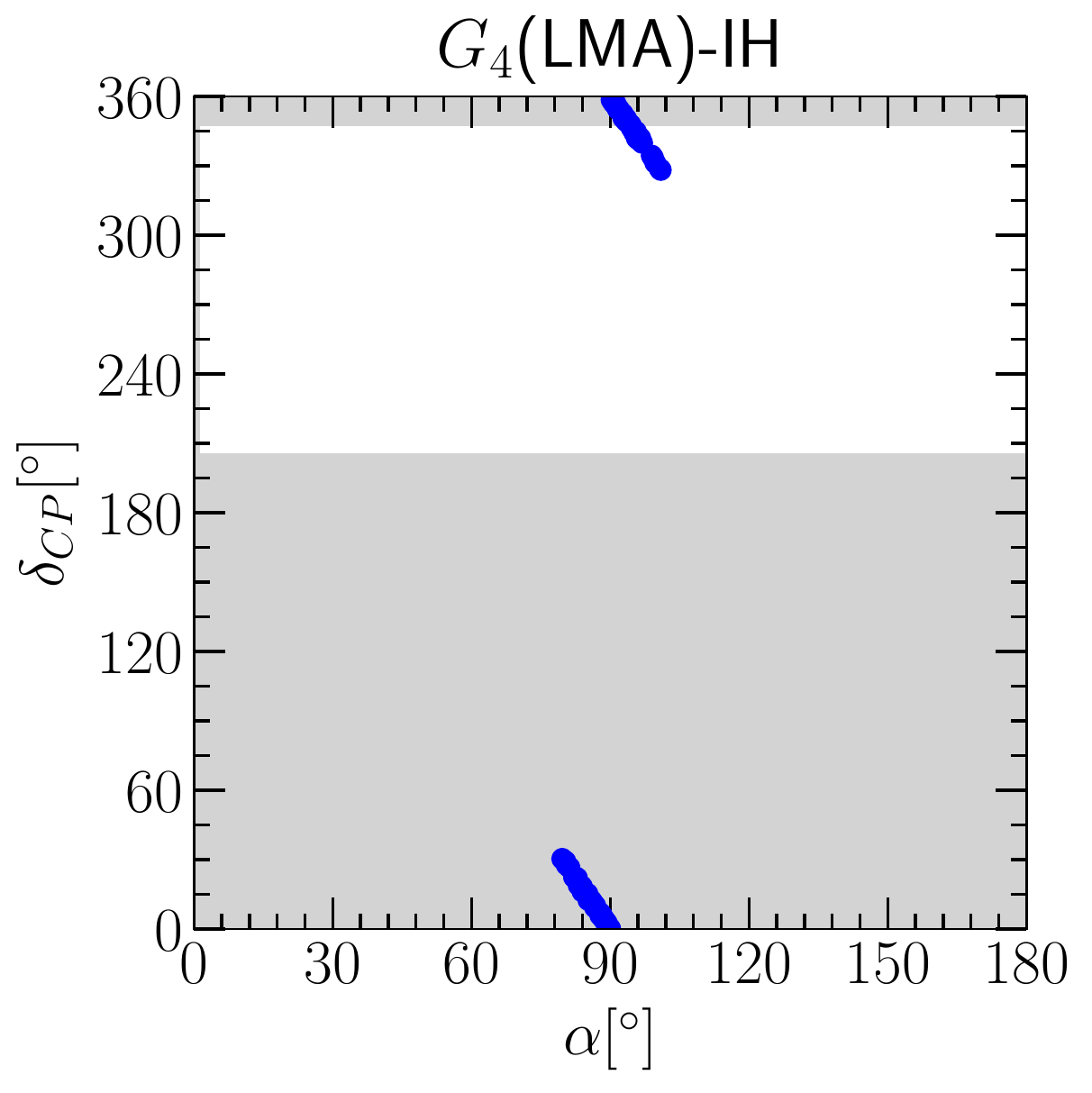}
\hspace{0.8 in}
\includegraphics[scale=0.41]{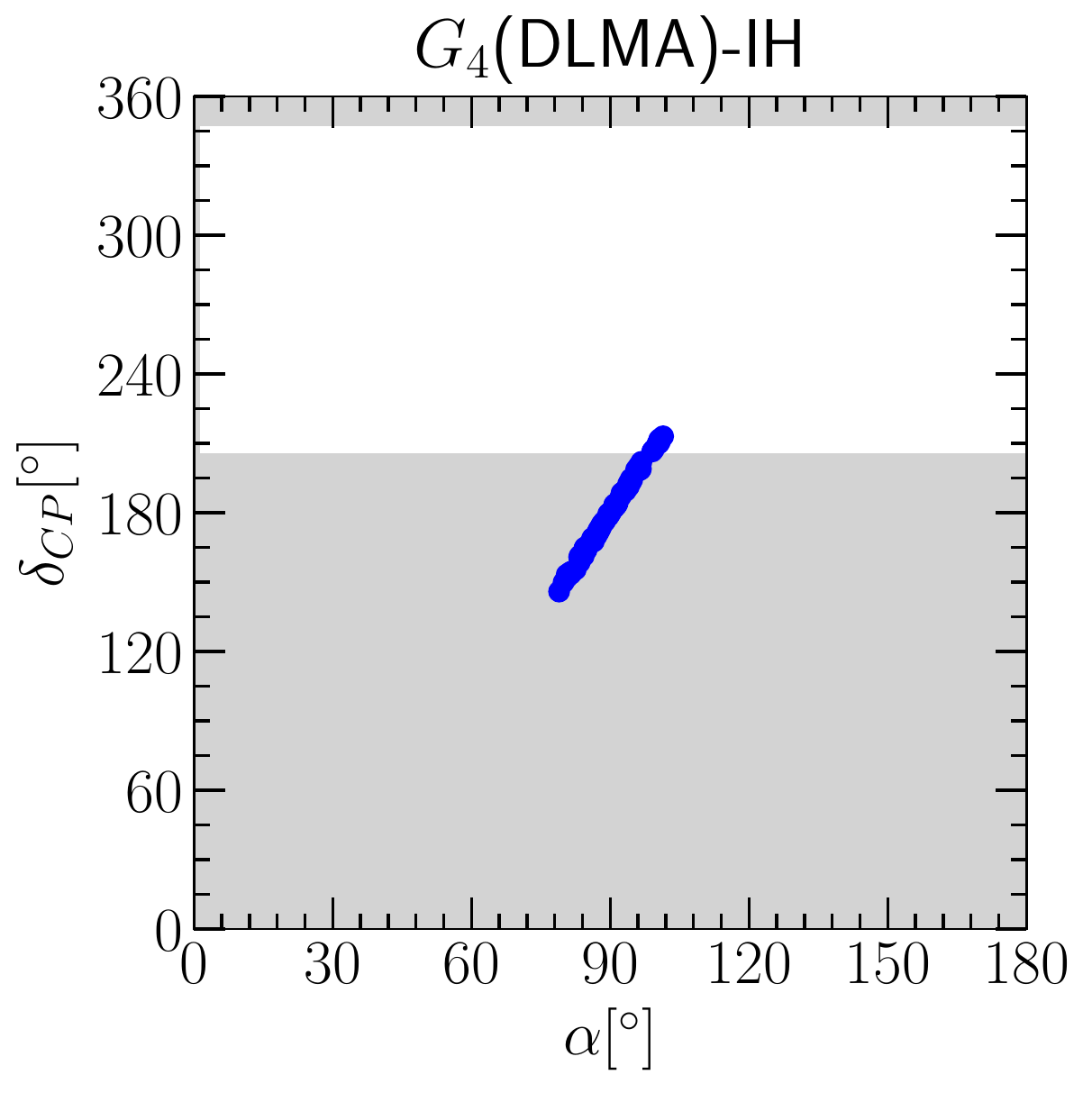}  \\
\includegraphics[scale=0.41]{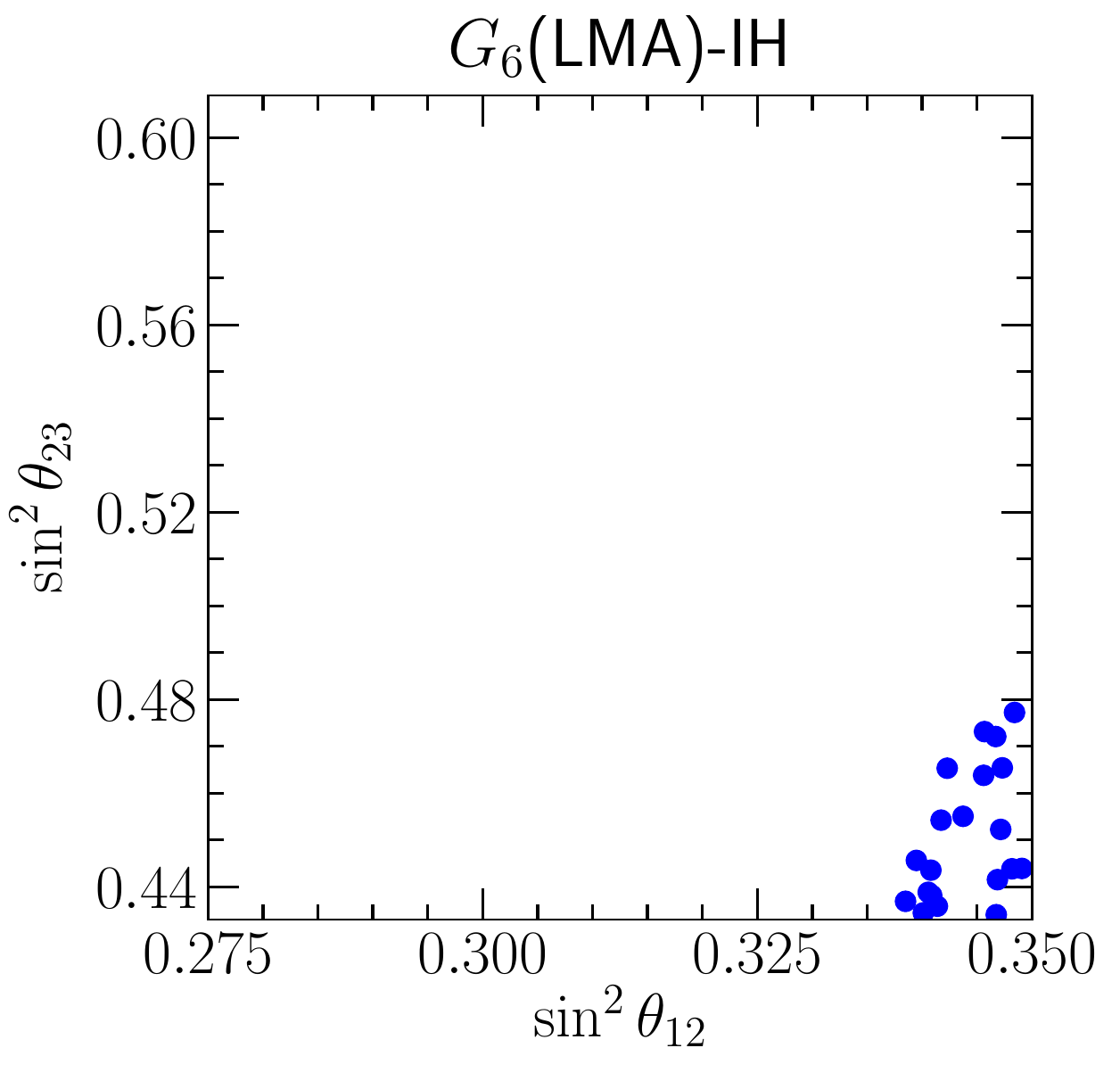} 
\hspace{0.8 in}
\includegraphics[scale=0.41]{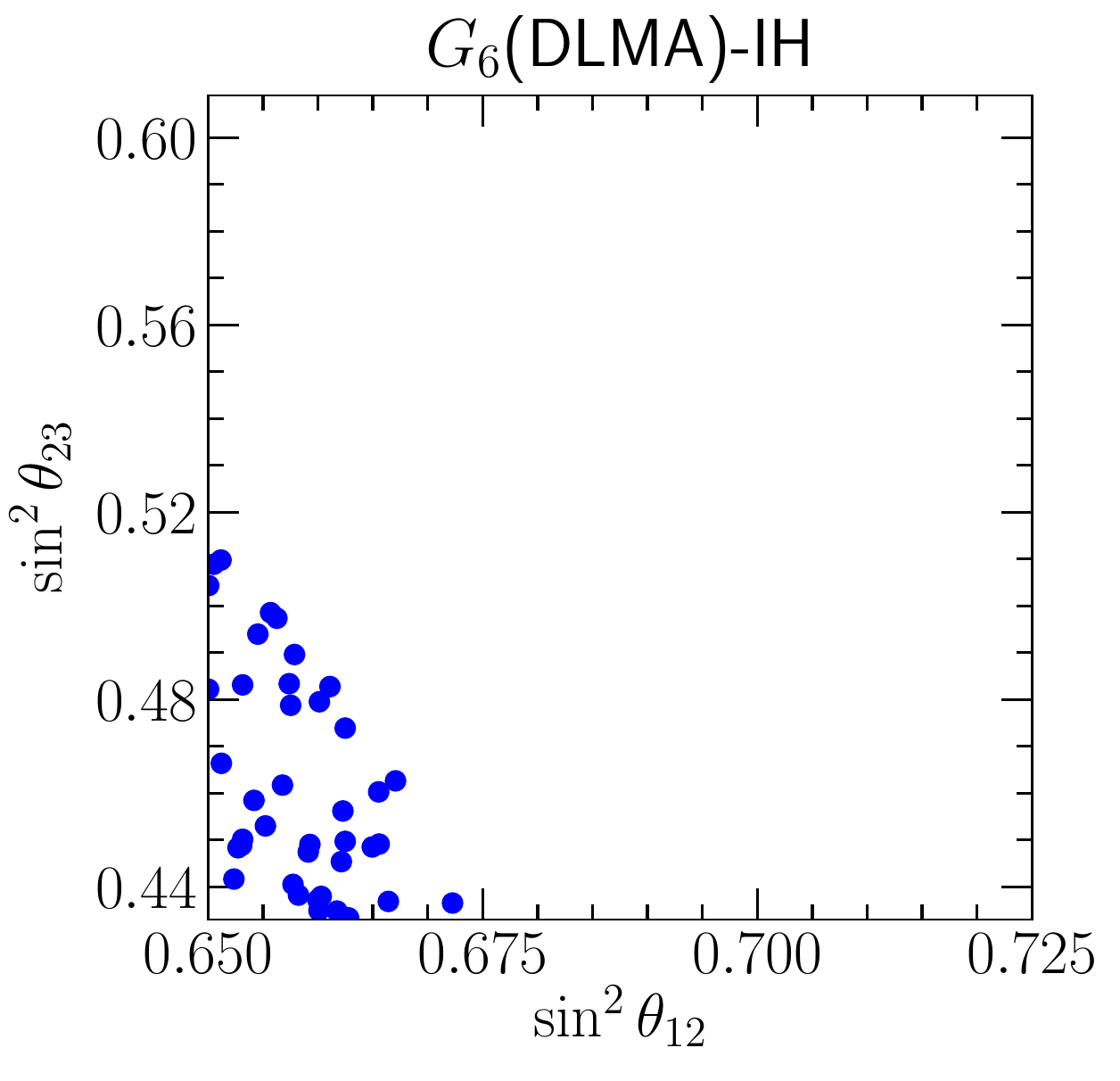}    
\caption{Correlations of the low energy parameters obtained for one zero texture classes. The grey shaded areas are the excluded values of $\delta_{\rm CP}$ at $3 \sigma$ as obtained by the global fit. }
\label{corr_T1}
\end{center}
\end{figure*}
One-zero texture implies that one of the elements of the low energy neutrino mass matrix is zero i.e.,
\begin{eqnarray}
M_{p q} = 0, 
\end{eqnarray}
where $p$, $q$ represent the three lepton flavors namely $e$, $\mu$ and $\tau$. 
We have summarized our results in Table \ref{one_zero} and presented the correlations between the various leptonic mixing parameters in Fig. \ref{corr_T1} for the allowed textures. Using the $3\sigma$ global fit values for the oscillation parameters along with the DLMA solution for the solar mixing angle as given in Table \ref{tabdata}, we have numerically checked for the parameter space which can lead to a vanishing matrix element corresponding to each class. 
Numerically, our definition of texture zero condition is given by $M_{p q} \leq10^{-8}$ with $p$, $q$ as $e$, $\mu$ and $\tau$. 
 Below we discuss each case in detail. 
\begin{table}[h!]
%\hspace{-1.186 in}
\renewcommand*{\arraystretch}{1.5}
\begin{center}
\begin{tabular}{|c|c|c|c|c|c|c|}
\hline
Texture & Element & LMA-NH & DLMA-NH & LMA-IH & DLMA-IH  \\
\hline
\hline
$G_1$ & $M_{ee} = 0$ &x & x & x & x  \\
\hline
$G_2$ & $ M_{e\mu} = 0$ & x & x & $\checkmark$ & $\checkmark$  \\
\hline
$G_3$ & $M_{e\tau} = 0$ & x & x & $\checkmark$  & $\checkmark$  \\
\hline
$G_4$ &$ M_{\mu\mu} = 0$ & x & x & $\checkmark$ & $$\checkmark$ $  \\
\hline
$G_5$ & $ M_{\mu\tau} = 0$ & x & x & x & x  \\
\hline
$G_6$ & $ M_{\tau\tau} = 0$ & x & x & $\checkmark$ & $\checkmark$  \\
\hline
\end{tabular}
\caption{Results of one-zero texture when the lowest neutrino mass is zero.}
\label{one_zero}
\end{center}
\end{table}
\begin{itemize}
\item With a vanishing $m_{\text{lowest}}$ none of the one zero texture case is allowed by the present data when the neutrino mass hierarchy is chosen to be normal. This can be understood in the following way. For NH, when the lowest mass i.e., $m_1$ is zero, then the texture zero condition requires cancellation between the $m_2$ and $m_3$ terms. As the difference between $m_2 \sim \Delta m^2_{\rm sol}$ and $m_3 \sim \Delta m^2_{\rm atm}$ terms are large, it is very difficult to achieve a cancellation between these two terms.  For inverted mass ordering the classes $G_1$ and $G_5$ are disfavored by the current oscillation data.  The textures $G_2$, $G_3$, $G_4$ and $G_6$ are allowed for IH and both for LMA and DLMA solutions of $\theta_{12}$. The textures $G_2$ and $G_4$ are related to textures $G_3$ and $G_6$ by $\mu-\tau$ symmetry respectively.  $\mu-\tau$ symmetry implies that $G_2$ ($G_4$) is related to $G_3$ ($G_6$) by the transformation $\theta_{23} \rightarrow 90^\circ - \theta_{23}$.

\item The textures $G_2$ and $G_3$ are allowed only for smaller values of $\sin\alpha$ with a nearly maximal $\sin\delta_{CP}$. Though we have shown this only for the LMA solution (first row of Fig. \ref{corr_T1}), the correlation is exactly similar for the DLMA case. 
We do not find any correlations among the mixing angles for these cases.
This can be understood from the analytical expression of $M_{e\mu}$ which is given by
\begin{eqnarray}
&\abs{M_{e \mu}} \simeq c_{13}\sqrt{\Delta m_{\text{atm}}^2} \abs [\Big] {c_{12} c_{23}s_{12} (e^{2 i \alpha}-1) - s_{13}s_{23}(c_{12}^2 + s_{12}^2 e^{2 i \alpha})e^{i \delta_{\rm CP}}}
\end{eqnarray}
To obtain the above expression, we have used $m_3 = 0$ and the approximations defined in Sec \ref{lowenergy}. From the above expression we understand that the texture $G_2$ will never be allowed if alpha is exactly $0^\circ$ or $180^\circ$. But small values of $\sin\alpha$ can make this element vanish. For small values of $\sin\alpha$, $M_{e \mu}$ is almost independent of $\theta_{12}$ and therefore it is allowed for all the values of $\theta_{12}$ including the LMA and the DLMA solutions.

\item For $G_4$, one needs to have the atmospheric mixing angle in the higher octant as realized from the second row of Fig. \ref{corr_T1}. It is also noticed that $M_{\mu \mu} =0$ is realized for higher values of $\theta_{12}$ of the allowed LMA range and lower values of $\theta_{12}$ of the allowed DLMA range. In addition to this for the class $G_4$, we see the restrictions on the CP phases with a large value of $\sin\alpha$ and small values for $\sin \delta_{\rm CP}$ (third row of Fig. \ref{corr_T1}). Here, we notice that the allowed values of $\delta_{CP}$ in $G_4$ are CP conserving and almost excluded by the current allowed values of $\delta_{CP}$ from the global data. The grey shaded areas in these plots are the excluded values of $\delta_{\rm CP}$ at $3 \sigma$ as obtained by the global fit. On the other hand, $G_4$ restricts $\alpha$ close to maximal, i.e., $90^\circ$. 
 For $G_6$, which is $\mu-\tau$ symmetric to $G_4$, we note that the lower octant for the atmospheric angle is allowed with a preference for the higher values of $\theta_{12}$ for the LMA solution whereas the smaller values of $\theta_{12}$ for the DLMA solution (fourth row of Fig. \ref{corr_T1}). The correlations between the Majorana phase $\alpha$ and Dirac phase $\delta_{CP}$ is similar as that of $G_4$.  

Let us try to understand the correlation between $\sin^2\theta_{12}$ and $\sin^2\theta_{23}$ of $G_4$ by analyzing the analytical expression of $M_{\mu\mu}$ which can be written as
\begin{eqnarray}
&\abs{M_{\mu \mu}} \simeq \sqrt{\Delta m_{\text{atm}}^2} \abs [\Big] {c_{23}^2 \left(s_{12}^2+c_{12}^2e^{2 i \alpha } \right)+\frac{1}{2} \left(1-e^{2 i \alpha }\right) \sin2\theta_{12} \sin2\theta_{23}  s_{13}e^{i \delta_{\rm CP}}}
\end{eqnarray}
In obtaining this expression, apart from using the same approximations which we used for $M_{e\mu}$, we have also omitted the terms of the order of $s_{13}^2$ and higher. For the texture $G_4$, this equation can be further simplified as 
\begin{eqnarray}
\theta_{12} =  0.5 \tan^{-1}\bigg(\pm \frac{c_{23}^2}{\sin2\theta_{23} s_{13}}\bigg)
\end{eqnarray}
where the `$+$' sign corresponds to the LMA solution i.e., $\alpha = 90^\circ$ and $\delta_{\rm CP} = 0^\circ$ and the `$-$' sign corresponds to the DLMA solution i.e., $\alpha = 90^\circ$ and $\delta_{\rm CP} = 180^\circ$. 
For $\theta_{23} =45^\circ$ and  $s_{13}^2 = 0.02$, we obtain $\sin^2\theta_{12} = 0.36 (0.64)$ for LMA (DLMA). 
Therefore, above equation is satisfied only if $\theta_{23}$ lies in the higher octant. Further,  from the above equation we understand that as $\theta_{23}$ increases from maximal value, $\theta_{12}$ decreases (increases) towards its allowed values for LMA (DLMA) and that is why for the texture $G_4$ to be allowed we need higher values of $\theta_{23}$ to allow the lower (higher) values of $\theta_{12}$ for the LMA (DLMA) solution.

Now let us briefly compare our results with the case when the lowest neutrino mass is non zero as obtained in Ref. \cite{Borgohain:2020now}. As we have already mentioned in the introduction, when the lowest neutrino mass is non zero, all the one-zero textures which are allowed with the LMA solution are also allowed with the DLMA solution except $G_1$. The later is allowed with the LMA solution in NH but disfavoured with the DLMA solution.  The textures $G_2$, $G_3$, $G_4$ and $G_6$ are allowed for both NH and IH, whereas the texture $G_5$ is allowed only in IH. However, we have seen that when the lowest neutrino mass is zero, only the textures $G_2$, $G_3$, $G_4$ and $G_6$ are allowed for IH and for both LMA and DLMA solutions.

\end{itemize}

\section{One zero texture and one vanishing minor}\label{TCS}

In this section we discuss the phenomenology of the neutrino mass matrix forms with the simultaneous appearance of one zero texture and one vanishing minor. Mathematically this condition can be expressed as \cite{Dev:2010if}
\begin{eqnarray}
M_{\nu(xy)} = 0 \\
M_{\nu(pq)} M_{\nu(rs)}  - M_{\nu(tu)} M_{\nu(vw)}  = 0.
\end{eqnarray}
These above two equations can be explicitly written as
\begin{eqnarray}\label{ozt}
m_1X +m_2 Ye^{-2 i \alpha}+m_3Ze^{2 i (-\beta+\delta)} =0 \\
m_1m_2A_3e^{-2 i \alpha} +m_2 m_3A_1e^{2 i \left(-\alpha-\beta+\delta\right)} +m_3m_1A_2e^{2 i \left(-\beta+\delta\right)} = 0
\end{eqnarray}
where, $X = U_{x1}U_{y1},Y = U_{x2}U_{y2},Z = U_{x3}U_{y3}$ and $A_h = \left(U_{pl}U_{ql}U_{rk}U_{sk} - U_{tl}U_{ul}U_{vk}U_{wk}\right) + (l \longleftrightarrow k)$, $h,l,k$ being the cyclic permutation of $(1,2,3)$. If we solve the above equations simultaneously, then we obtain
\begin{align}
 &\frac{m_1}{m_3}e^{2 i \beta} = -\frac{\left(XA_1- YA_2+ZA_3 \pm \sqrt{X^2A_1^2+(YA_2-ZA_3)^2 - 2 XA_1(YA_2+ZA_3)} \right)}{2 X A_3}e^{2 i \delta} \label{mratio1}\\
  &\frac{m_1}{m_2}e^{2 i \alpha} = \frac{\left(-XA_1- YA_2+ZA_3 \pm \sqrt{X^2A_1^2+(YA_2-ZA_3)^2 - 2 XA_1(YA_2+ZA_3)} \right)}{2 X A_2}\label{mratio2}
 \end{align}
Thus there can be four possible solutions of the mass ratios which can lead to a one zero texture and vanishing minor simultaneously. We will denote them by $(+,+),(+,-),(-,+),(-,-)$. If we define,
\begin{align}
&S = -\frac{\left(XA_1- YA_2+ZA_3 \pm \sqrt{X^2A_1^2+(YA_2-ZA_3)^2 - 2 XA_1(YA_2+ZA_3)} \right)}{2 X A_3}e^{2 i \delta}\\
 &T = \frac{\left(-XA_1- YA_2+ZA_3 \pm \sqrt{X^2A_1^2+(YA_2-ZA_3)^2 - 2 XA_1(YA_2+ZA_3)} \right)}{2 X A_2},
\end{align}
then the mass ratios can be determined as
\begin{align}
&\rho = \frac{m_1}{m_3} = \left|S\right| \\
&\sigma = \frac{m_1}{m_2} = \left| T \right|,
\end{align}
and the Majorana phases can be predicted as 
\begin{align}
&\alpha =  \frac{1}{2} arg\left(T\right) \\
&\beta = \frac{1}{2}arg \left( S\right).
\end{align}
Further using $\rho$ and $\sigma$ we can construct two independent expressions of $m_1$ in the following way
\begin{align}
&m_1 = \rho \sqrt{\frac{\Delta m^2_{21}+\Delta m^2_{31}}{1-\rho^2}} \\
&m_1 = \sigma \sqrt{\frac{\Delta m^2_{21}}{1-\sigma^2}}. 
\end{align} 
The condition of simultaneous appearance of one texture zero and vanishing minor will only be satisfied when the values of $m_1$ calculated independently from $\rho$ and $\sigma$ are equal. Numerically, when this condition is satisfied, we obtain $M_{\nu(xy)} \leq 10^{-8}$ and $M_{\nu(pq)} M_{\nu(rs)}  - M_{\nu(tu)} M_{\nu(vw)}  \leq 10^{-8}$ simultaneously with $p$, $q$ as $e$, $\mu$ and $\tau$.

\begin{table}[h]
%\hspace{-0.5 in} 
\begin{center}
\begin{small}
\begin{tabular}{|c|c|c|c|c|c|c|c|}
\hline  &  & A. & B. &C. & D. & E.& F. \\
\hline 1. & $\left(
\begin{array}{ccc}
0 & b & c \\  b & d & e \\ c& e & f
\end{array}
\right)$ & $df - e^{2}=0$ & $bf-ec=0$ & $be-cd=0$ &  Two Zero &  Two Zero&  Two Zero \\
\hline 2. &$\left(
\begin{array}{ccc}
a & 0 & c \\  0& d & e \\ c& e & f
\end{array}
\right)$   & $df-e^2=0$ &Two Zero &Two Zero  & $af-c^2=0$ & Two Zero&Two Zero \\
\hline 3. & $\left(
\begin{array}{ccc}
a & b & 0 \\ b & d & e \\0 &e  & f
\end{array}
\right)$ &$df-e^2=0$  & Two Zero & Two Zero & Two Zero &Two Zero  & $ad-b^2=0$\\
\hline 4. & $\left(
\begin{array}{ccc}
a & b & c \\ b & 0 & e \\c & e & f
\end{array}
\right)$ & Two Zero &$bf-ec=0$  & Two Zero &$af-c^2=0 $ & $ae-bc=0$&Two Zero \\
\hline 5. &$\left(
\begin{array}{ccc}
a & b & c \\  b & d & 0 \\ c & 0 & f
\end{array}
\right)$  & Two Zero &Two Zero  & Two Zero & $af-c^2=0$ & Two Zero &$ad-b^2=0$\\
\hline 6. & $\left(
\begin{array}{ccc}
a & b & c \\ b & d & e \\ c& e & 0
\end{array}
\right)$ & Two Zero &Two Zero  & $be-dc=0$ & Two Zero &$ae-bc=0$  &$ad-b^2=0$\\
\hline
\end{tabular}
\end{small}
\caption{36 texture structures of $M_{\nu}$ with a texture Zero and one vanishing minor.}
\label{tc_fits}
\end{center}
\end{table}
\begin{table}[h!]
%\hspace{-1.186 in}
\renewcommand*{\arraystretch}{1.5}
\begin{center}
\begin{tabular}{|c|c|c|c|c|c|}
\hline
Texture & LMA-NH & DLMA-NH & LMA-IH & DLMA-IH  \\
\hline
\hline
$A_1$ & x & x & x & x  \\
\hline
$B_1$ &  x & x & x & x  \\
\hline
$C_1$ & x & x & x  & x  \\
\hline
\hline
$A_2$ & x & $\checkmark$ (+,+) & $\checkmark$ (-,-) & $\checkmark$ (-,-)  \\
\hline
$D_2$ & $\checkmark$ (-,-), (+,+) & $\checkmark$ (-,-), (+,+) & $\checkmark$ (-,-) & $\checkmark$ (-,-)  \\
\hline
\hline
$A_3$ & x & $\checkmark$ (-,-) & $\checkmark$ (+,+) & $\checkmark$ (+,+)  \\
\hline
$F_3$ &  $\checkmark$ (-,-), (+,+) &  $\checkmark$ (-,-), (+,+) &  $\checkmark$ (+,+) &  $\checkmark$ (+,+)  \\
\hline
\hline
$B_4$ & $\checkmark$ (-,-) & $\checkmark$ (-,-) & $\checkmark$ (-,-), (+,+) & $\checkmark$ (-,-), (+,+)  \\
\hline
$D_4$ &  x & x & x & x  \\
\hline
$E_4$ & x & x & x  & x  \\
\hline
\hline
$D_5$ & x & x & x & x  \\
\hline
$F_5$ & x & x & x & x  \\
\hline
\hline
$C_6$ & $\checkmark$ (-,-) & $\checkmark$ (-,-) & $\checkmark$ (-,-), (+,+) & $\checkmark$ (-,-), (+,+)  \\
\hline
$E_6$ &  x & x & x & x  \\
\hline
$F_6$ & x & x & x  & x  \\
\hline
\end{tabular}
\caption{Results of one zero texture with one vanishing minor when lowest mass is non-zero. The allowed textures are marked with $\checkmark$ and which is true only for the solutions of Eq. \ref{mratio1} and Eq. \ref{mratio2} mentioned in parentheses.}
\label{v_minor}
\end{center}
\end{table}
As tabulated in Table \ref{tc_fits} there are 36 possible cases which can allow the neutrino mass matrix to have one zero texture and a vanishing minor simultaneously. Among them certain cases belong to the two zero texture classes. Thus we have a total of 15 distinct cases with one texture zero and one vanishing minor that are not reducible to a two zero texture case.

In the Table \ref{v_minor} we summarize the validity of each class of this particular scheme of neutrino mass matrix taking LMA as well as DLMA solution for both the neutrino mass orderings and presented the correlations between different observables in Figs. \ref{corr_TC1},  \ref{corr_TC2}, and \ref{corr_TC3} when the lowest neutrino mass is non-zero\footnote{We have checked that when the lowest neutrino mass is zero, none of the cases are allowed.}.
 
From Table \ref{v_minor} we understand that all the cases which are allowed for the LMA solution are also allowed for the DLMA solution except $A_2$ and $A_3$. $A_2$ and $A_3$ are the two cases which are allowed in DLMA with NH but not allowed for LMA with NH.

From Table \ref{v_minor} we also see that some of the cases are allowed only for some specific solutions of the mass ratios but not for all the four solutions. The classes $A_1$, $B_1$, $C_1$, $D_4$, $E_4$, $D_5$, $F_5$, $E_6$ and $F_6$ are not allowed by the recent data irrespective of the choice of the solution for the solar mixing angle.  
From the correlation plots we note the following:

\begin{figure*}[h!]
\begin{center}
\includegraphics[scale=0.41]{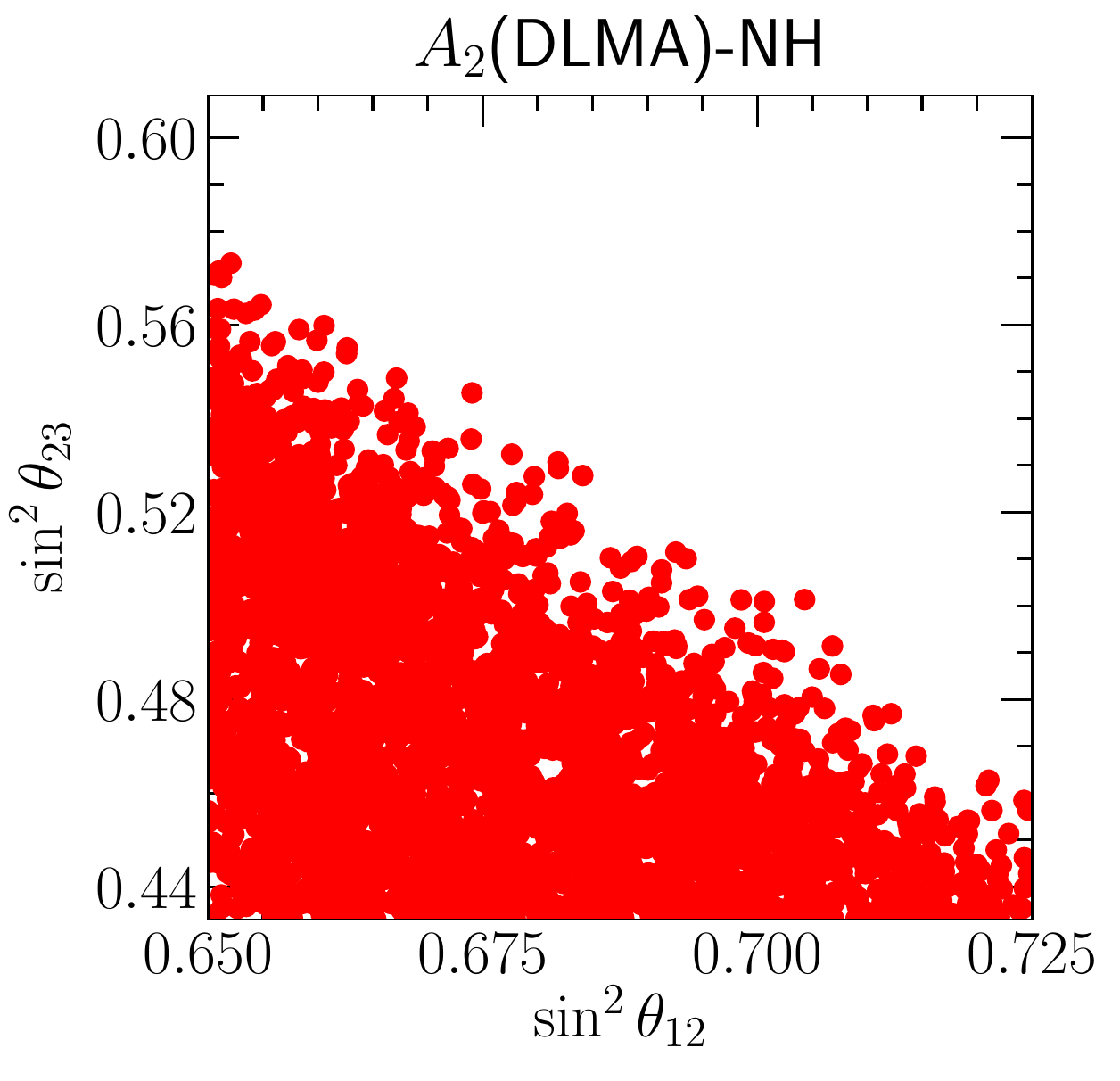}
\includegraphics[scale=0.41]{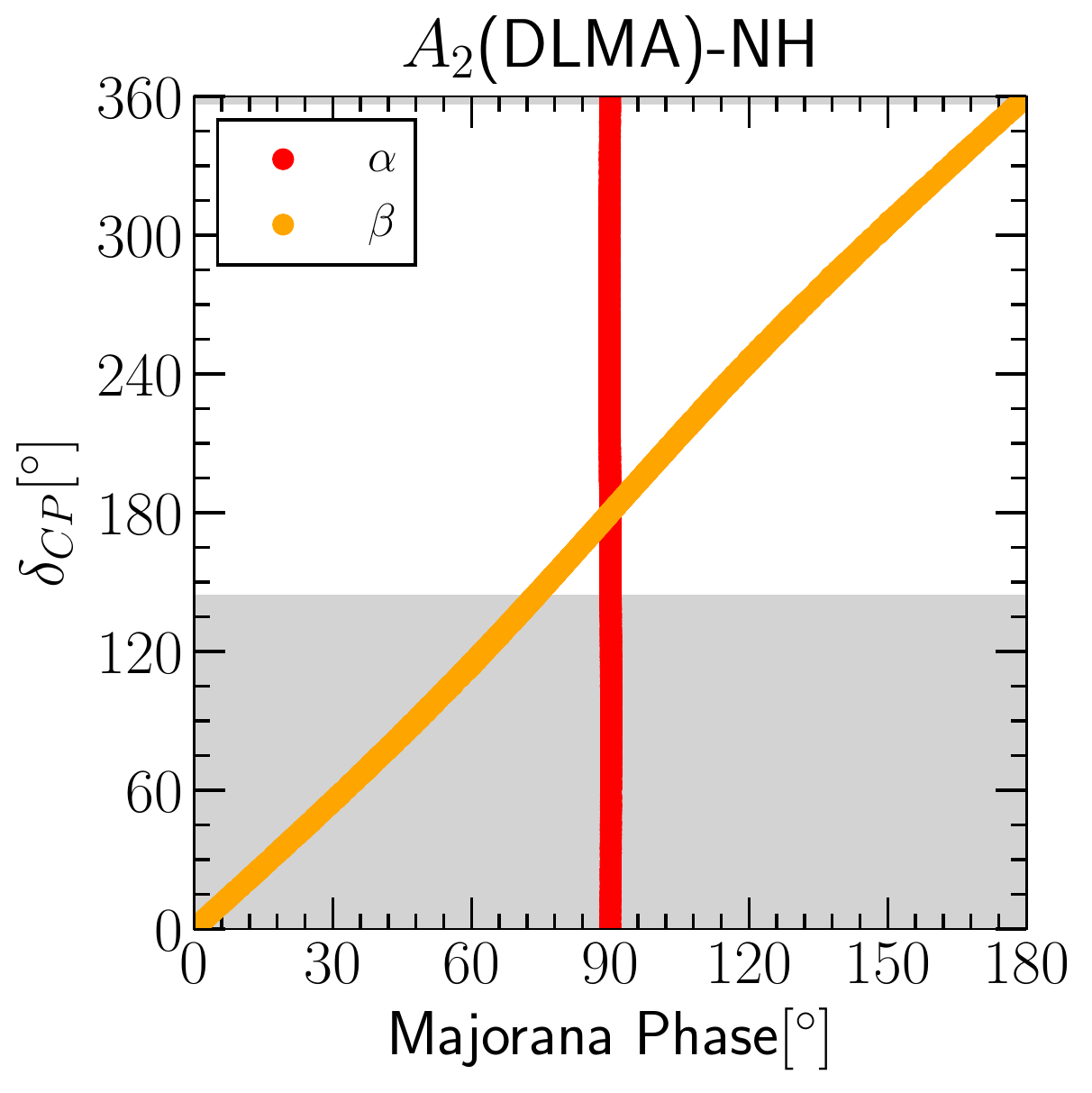}  
\includegraphics[scale=0.41]{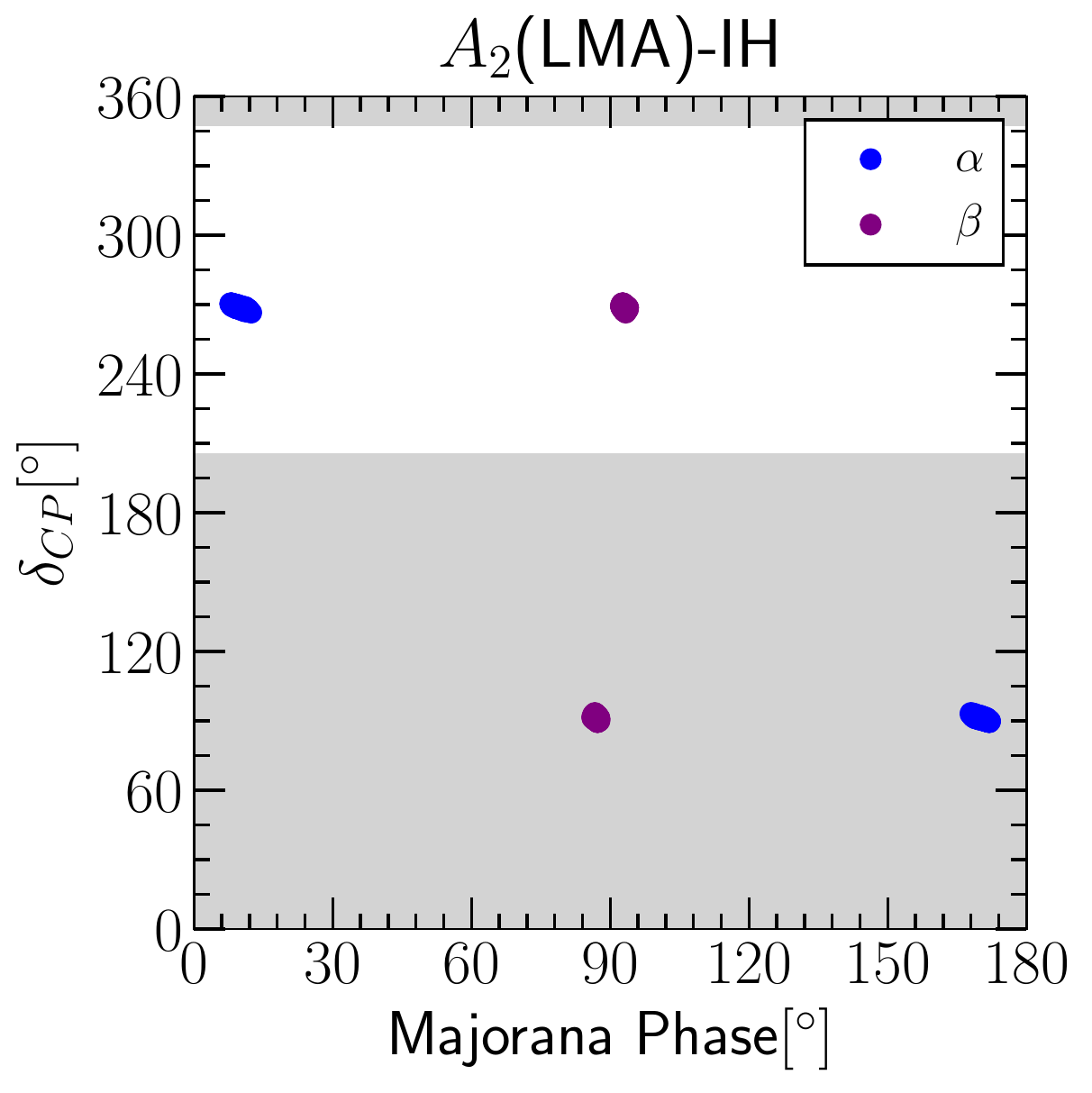} \\
\includegraphics[scale=0.41]{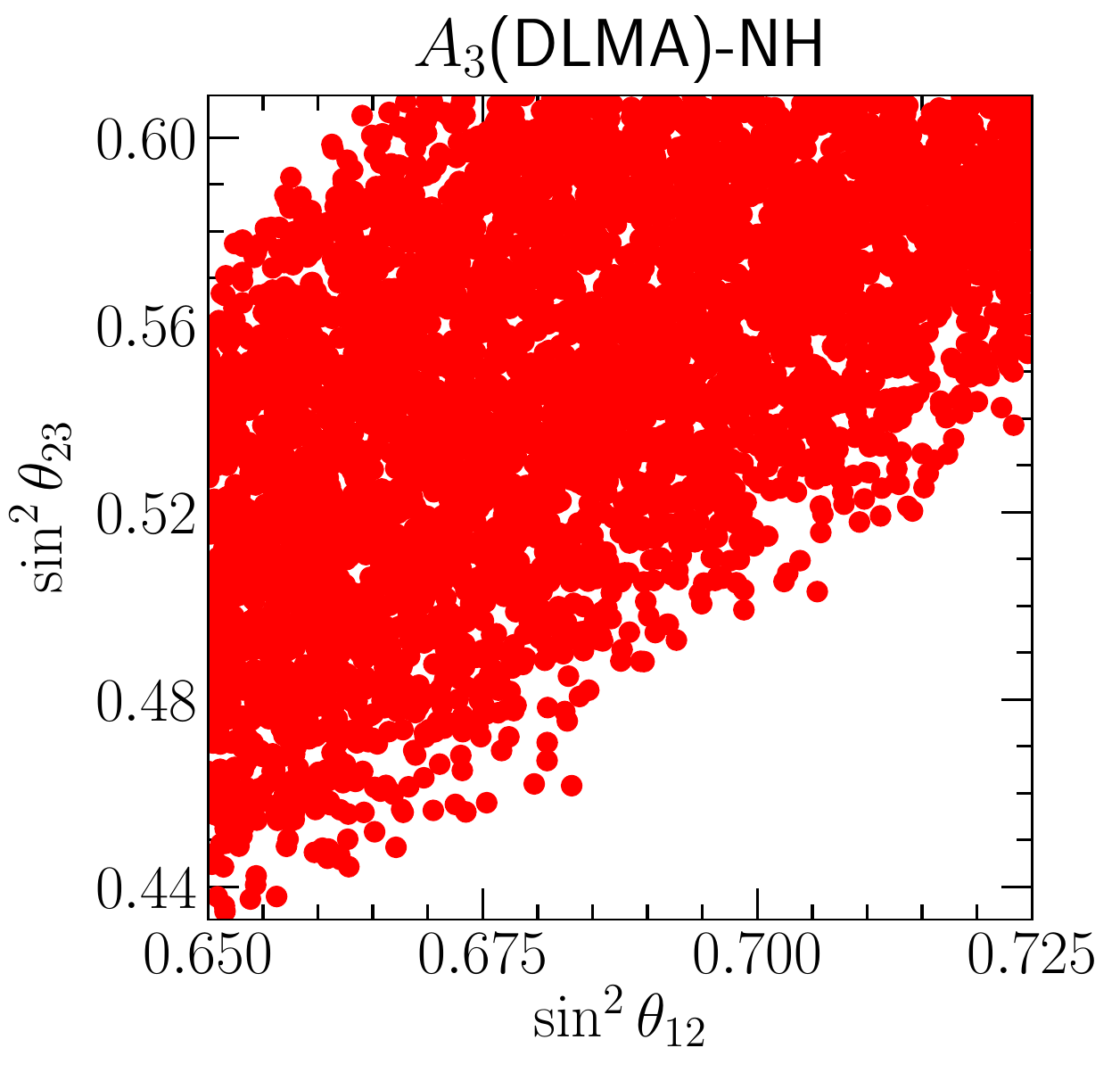}
\includegraphics[scale=0.41]{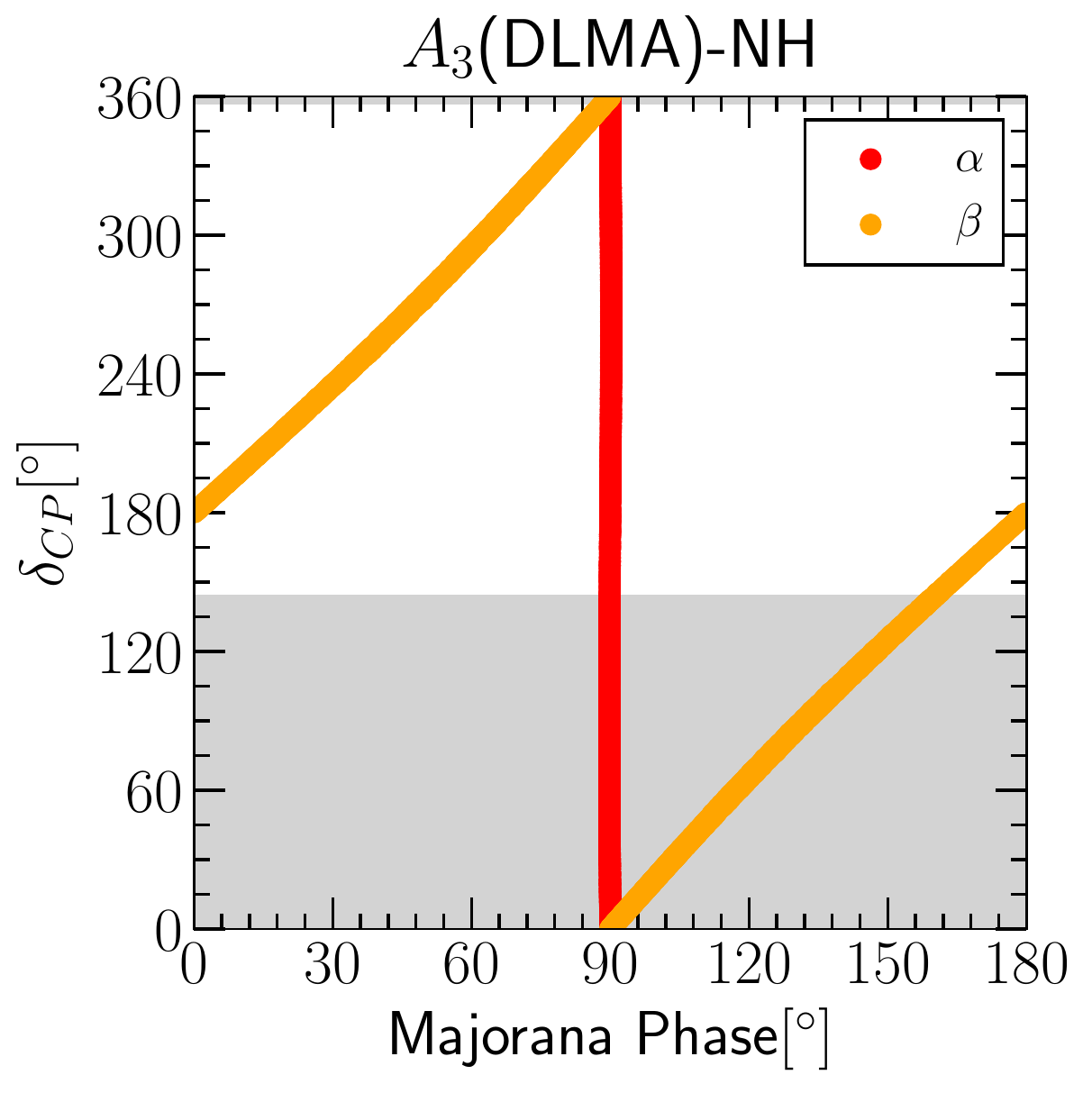} 
\includegraphics[scale=0.41]{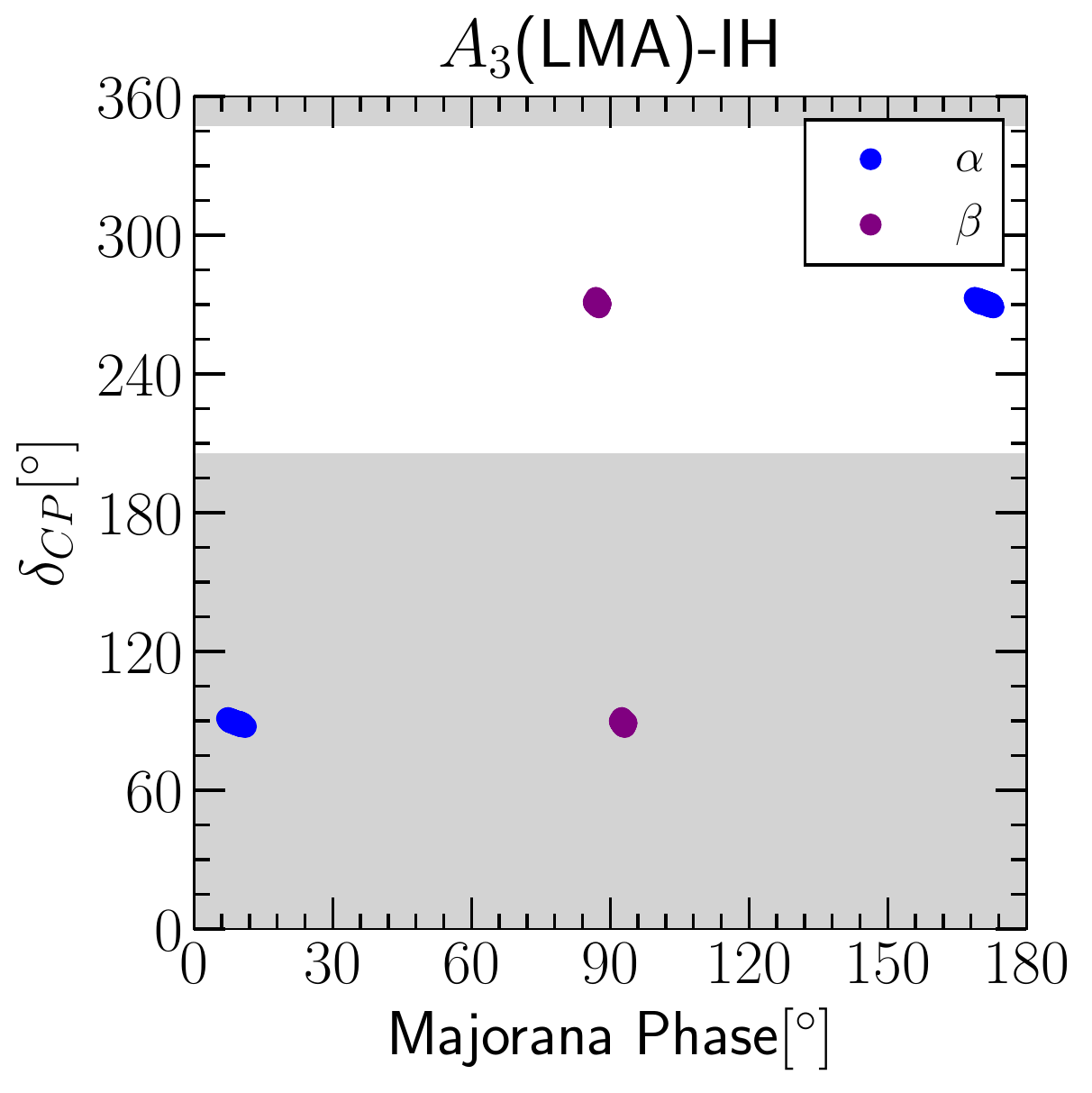} 
\caption{Correlation plots for the classes $A_2$ and $A_3$. For explanation please see the text. The grey shaded areas in these plots are the excluded values of $\delta_{\rm CP}$ at $3 \sigma$ as obtained by the global fit.}
\label{corr_TC1}
\end{center}
\end{figure*}

\begin{itemize}
\item In the Fig. \ref{corr_TC1} we present the possible correlations among the neutrino parameters for the classes $A_2$ and $A_3$ which are related by $\mu-\tau$ symmetry. 
For the class $A_2$ with NH, the correlation between the atmospheric angle and the DLMA solution for the solar angle shows that for the smaller values of $
\theta_{12}$ for the DLMA solution we can have both the octants for $\theta_{23}$, while the larger DLMA values are likely to favor the lower octant of  $\theta_{23}$ (top left panel). For the $A_3$ one can notice that the smaller DLMA values prefer both the octants for $\theta_{23}$, whereas the larger DLMA values prefer higher octant of $\theta_{23}$ in NH (bottom left panel). For the exact cancellations to make the $A_2$ and $A_3$ class allowed for NH, one needs a highly constrained value for the Majorana phase $\alpha$ which is around $\pi/2$, whereas the remaining two phases $\delta_{\rm CP}$ and $\beta$ can have values in the range $0 - 2\pi$ and $0-\pi$ respectively (top middle and bottom middle panels). For IH the correlations among the CP phases are found to be completely distinct as compared to the NH case. The correlations obtained among the three CP phases for $A_2$ and $A_3$ ensure that we need a nearly maximal $\sin\delta_{\rm CP}$ and $\sin\beta$ and small $\sin\alpha$ values for the LMA solutions (top right and bottom right panels). For the DLMA solution in IH, the correlations are similar as that of LMA solutions. Now let us try to understand the above described results for $A_2$ from the following analytical expressions. Using the $(+,+)$ solution for $A_2$ we obtain
\begin{gather}
S = \frac{c_{12} s_{13}s_{23} e^{i \delta_{\rm CP}}}{s_{12}c_{23}},~T = -\frac{c_{12}^2}{s_{12}^2}
\end{gather}
which implies
\begin{gather}
\frac{m_1}{m_3} = \frac{c_{12} s_{13}s_{23}}{s_{12}c_{23}},~\frac{m_1}{m_2} = \frac{c_{12}^2}{s_{12}^2} \\
\alpha = 90^\circ,~\beta = \frac{\delta_{\rm CP}}{2}.
\end{gather}
From the above expression we understand that the mass ratio $\frac{m_1}{m_2}$ can only be less than one for the DLMA solution of $\theta_{12}$ and the mass ratio $\frac{m_1}{m_3}$ is always less than one, implying NH. Therefore the (+,+) solution is allowed only for NH and DLMA solution of $\theta_{12}$. It is also important to note that the the prediction for the Majorana phases and the correlation of the Dirac phase $\delta_{CP}$ as obtained in the numerical analysis is correctly reproduced from the analytical expressions. Let us now see the same for the $(-,-)$ solution of $A_2$. For $(-,-)$ solution we obtain
\begin{gather} 
S = - \frac{1}{s_{13}^2} + \mathcal{O} [\frac{1}{s_{13}}],~~~ T =  1-\frac{e^{i \delta } s_{13} s_{23}}{c_{12} c_{23} s_{12}}
\end{gather} 
which implies
\begin{gather}
\frac{m_1}{m_3} =   \frac{1}{s_{13}^2},~\frac{m_1}{m_2} =1-\frac{2 \cos\delta_{\rm CP} s_{13} s_{23}}{c_{12} c_{23} s_{12}} \\
\alpha = \frac{1}{2}\tan^{-1}\left(\frac{s_{13}s_{23} \sin\delta_{\rm CP}}{c_{12}c_{23}s_{12} - s_{13}s_{23}\cos\delta_{\rm CP}}\right),~\beta = 90^\circ.
\end{gather}
From the above equations we note that the mass ratio $\frac{m_1}{m_3}$ is always greater than one implying IH. Further we notice that to have $\frac{m_1}{m_2}$ less than one, we need positive values of $\cos\delta_{CP}$ for both LMA and DLMA solution of $\theta_{12}$. Therefore we understand that for $(-,-)$ solution $A_2$ is allowed in IH for both LMA and DLMA solution. Regarding the values of the phases, from the numerical simulation we obtain $\delta_{\rm CP}$ close to $90^\circ$ and $270^\circ$ and this gives $m_1/m_2 < 1$. 
We also obtain the value of  $\beta$ as equals to $90^\circ$ which is reproduced by our analytical expression. For $\alpha$, from the analytic expressions we obtain 
\begin{eqnarray}
\alpha = \frac{1}{2}\tan^{-1}\left(\pm \frac{s_{13}}{c_{12}s_{12}}\right)
\end{eqnarray}
where we have used $\theta_{23} = 45^\circ$. The $+ (-)$ sign is for $\delta_{CP} = 90^\circ (270^\circ$). Further using the best-fit values of $\theta_{12} = 33^\circ$ or $57^\circ$ and $\theta_{13} = 8.5^\circ$ we obtain, $\alpha = 9^\circ (171^\circ)$ for $\delta_{CP} = 90^\circ (270^\circ)$. This correctly matches with our numerical results as presented in Fig. \ref{corr_TC1}.

\item In Fig. \ref{corr_TC2} we present the correlation plots for $D_2$ and $F_3$ which are again $\mu-\tau$ symmetric. First let us discuss the case for NH. For $D_2$ ($F_3$), the $(+,+)$ solution corresponds to the region $m_1 < (>) 0.02$  eV and the $(-,-)$ solution corresponds to the region $m_1 > (<) 0.02$ for both LMA and DLMA solution.  For $D_2$ ($F_3$), all the values of $\theta_{23}$ are allowed for $(+,+)$ ($(-,-)$) solution and only higher (lower) octant is allowed for $(-,-)$ ($(+,+)$) solution in LMA. However for DLMA only lower (higher) octant is allowed for $D_2$ ($F_3$). 
 
For NH both the LMA and DLMA show analogous predictions on the Majorana phases for the choice of $(-,-)$ ($(+,+)$) for $D_2$ ($F_3$), but the other choice of solution gives rise to a considerably different region of parameter space for all the CP phases. One can notice that for the solution $(+,+)$ ($(-,-)$), $D_2$ ($F_3$) for LMA-NH needs a larger $\sin\alpha$ and $\sin\delta$ with a smaller $\sin\beta$, but for DLMA-NH the one needs to have larger and almost similar range of values for $\sin \alpha$ and $\sin \beta$ for relatively smaller values for $\sin\delta$.
These constraints on the $m_{\text{lowest}}$ and the Majorana phases can be insightful in the study of effective neutrino mass $m_{\beta\beta}$ governing the $0\nu\beta\beta$ process. 
It is seen that $D_2$ ($F_3$) with IH can survive for both the LMA and DLMA choices with a strict preference of lower (higher) octant for the  $\theta_{23}$. Here for both the class $D_2$ and $F_3$ with IH, all the three CP phases are tightly constrained to very narrow regions with $\delta_{CP} \sim 90^\circ$ and the Majorana phases around $0^\circ$ and $180^\circ$. 

\begin{figure*}[h!]
\begin{center}
\includegraphics[scale=0.37]{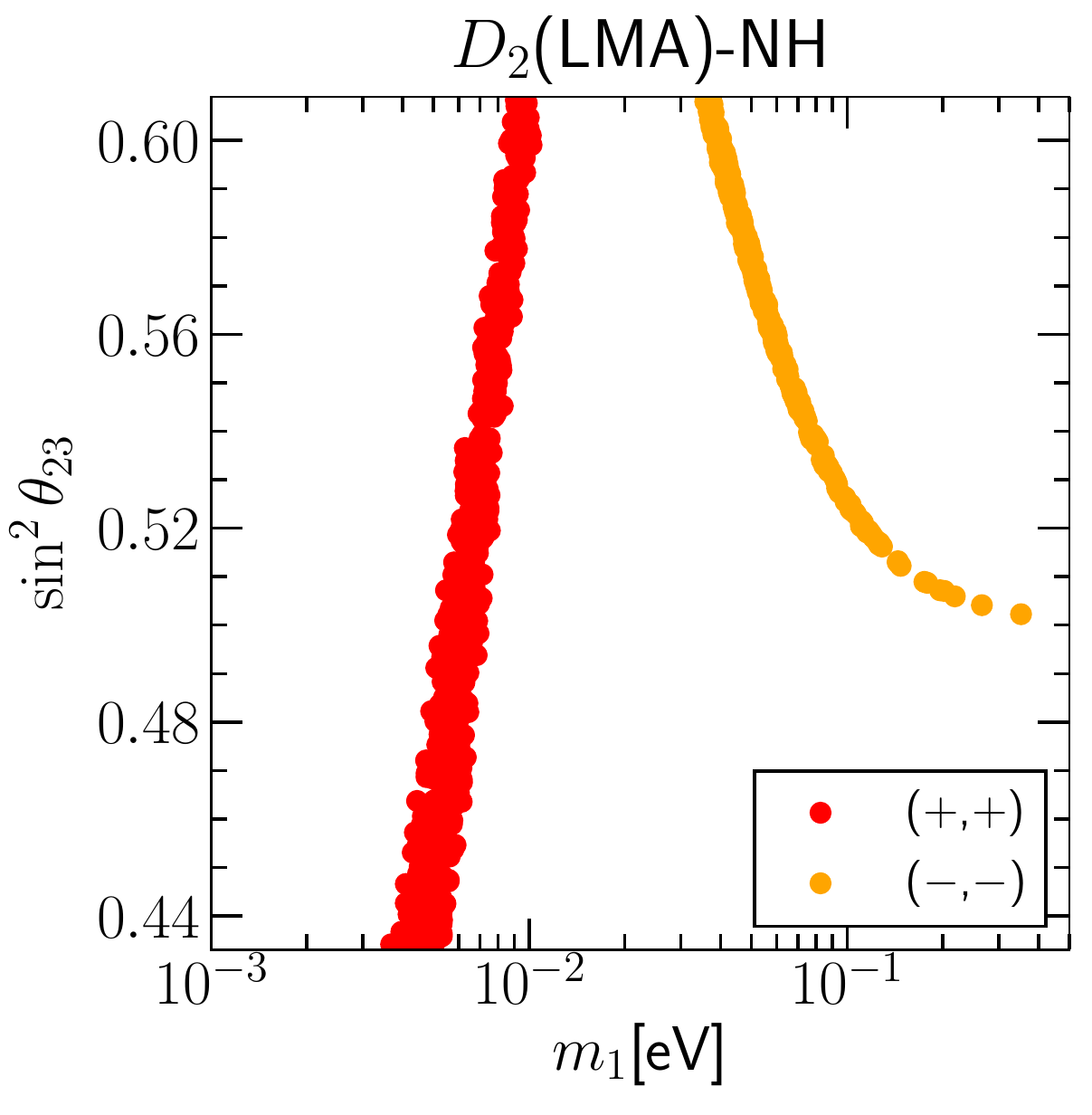}
\includegraphics[scale=0.37]{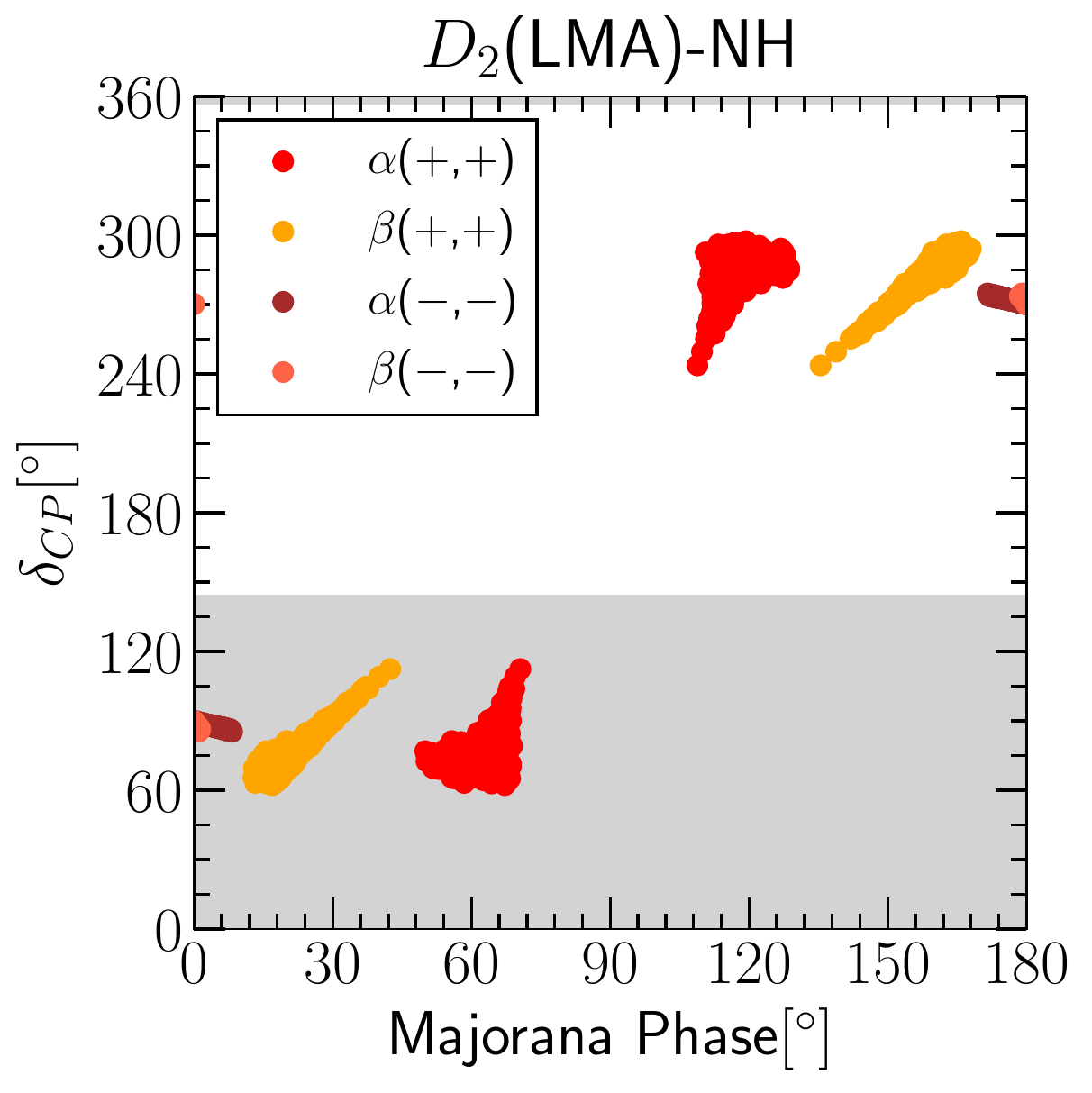}   
\includegraphics[scale=0.37]{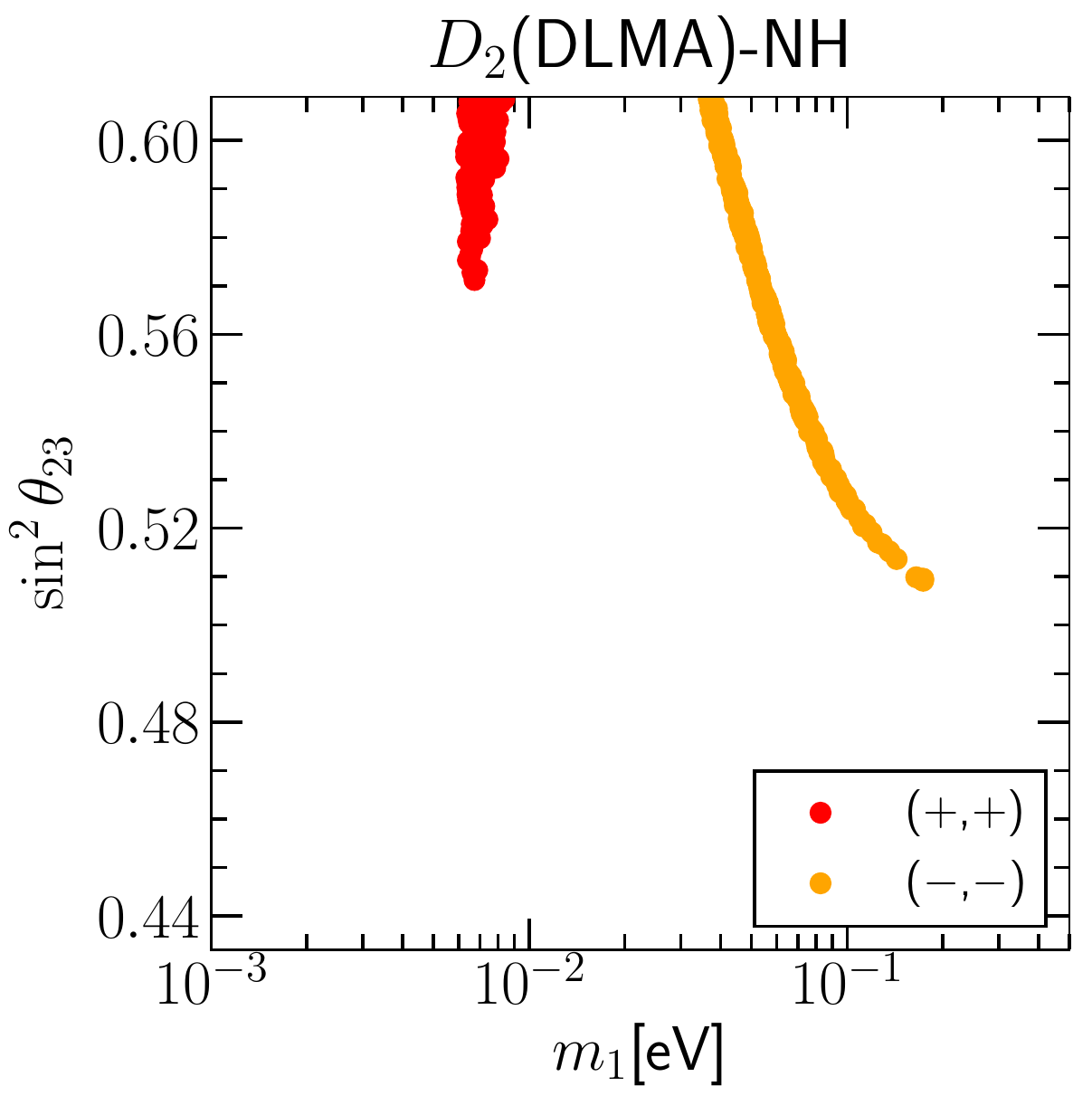} \\
\includegraphics[scale=0.37]{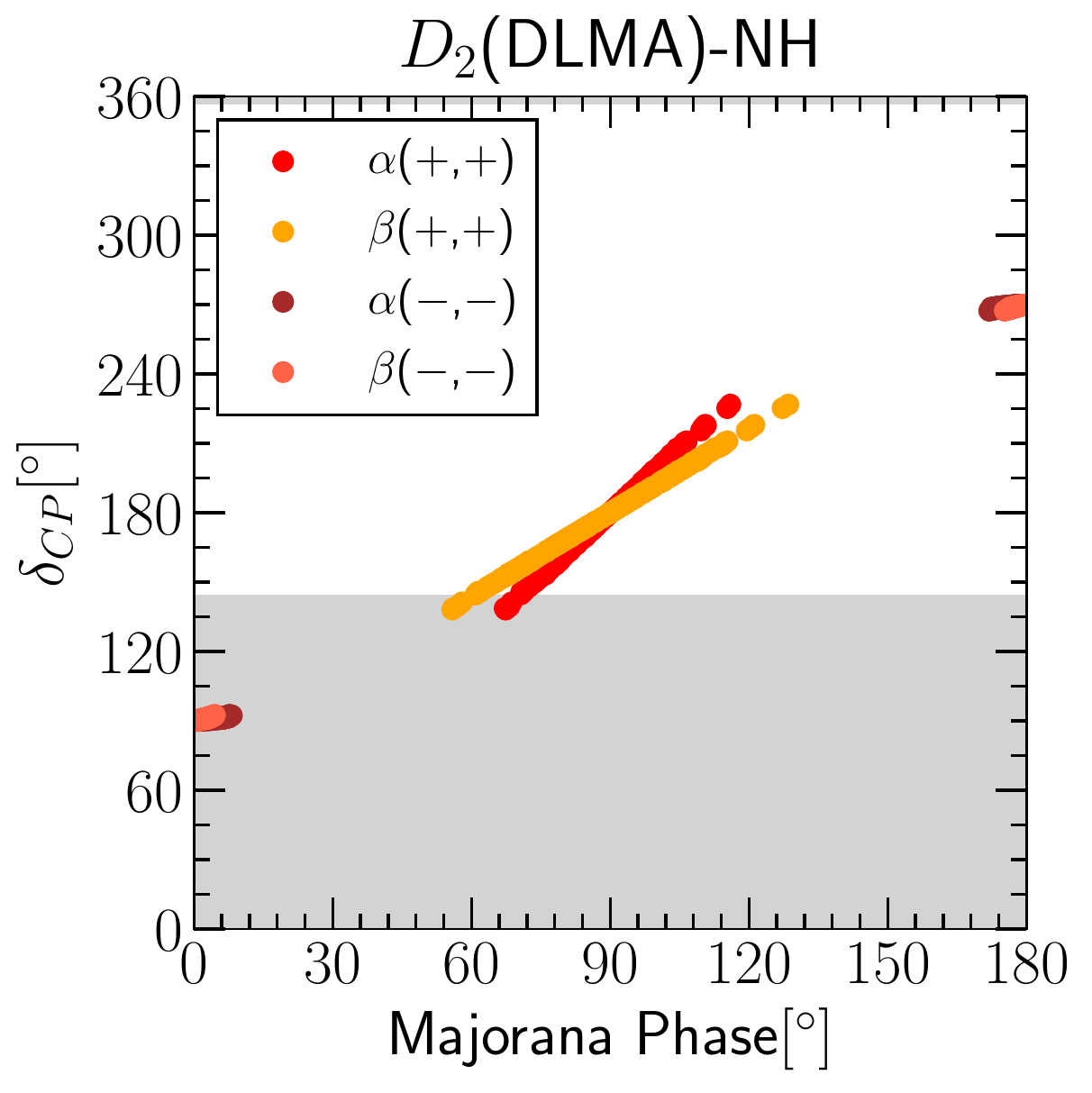}      
\includegraphics[scale=0.37]{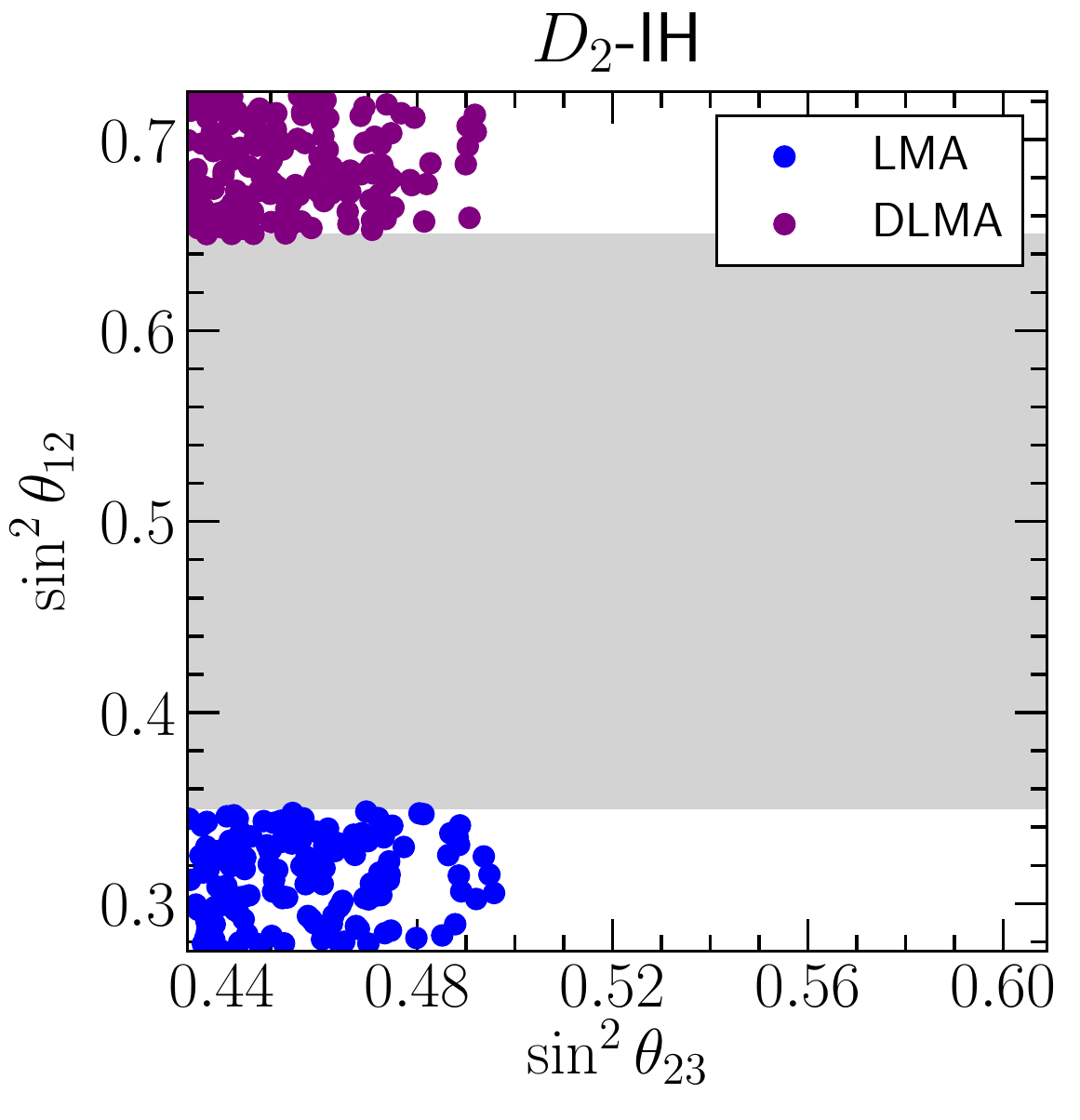}
\includegraphics[scale=0.37]{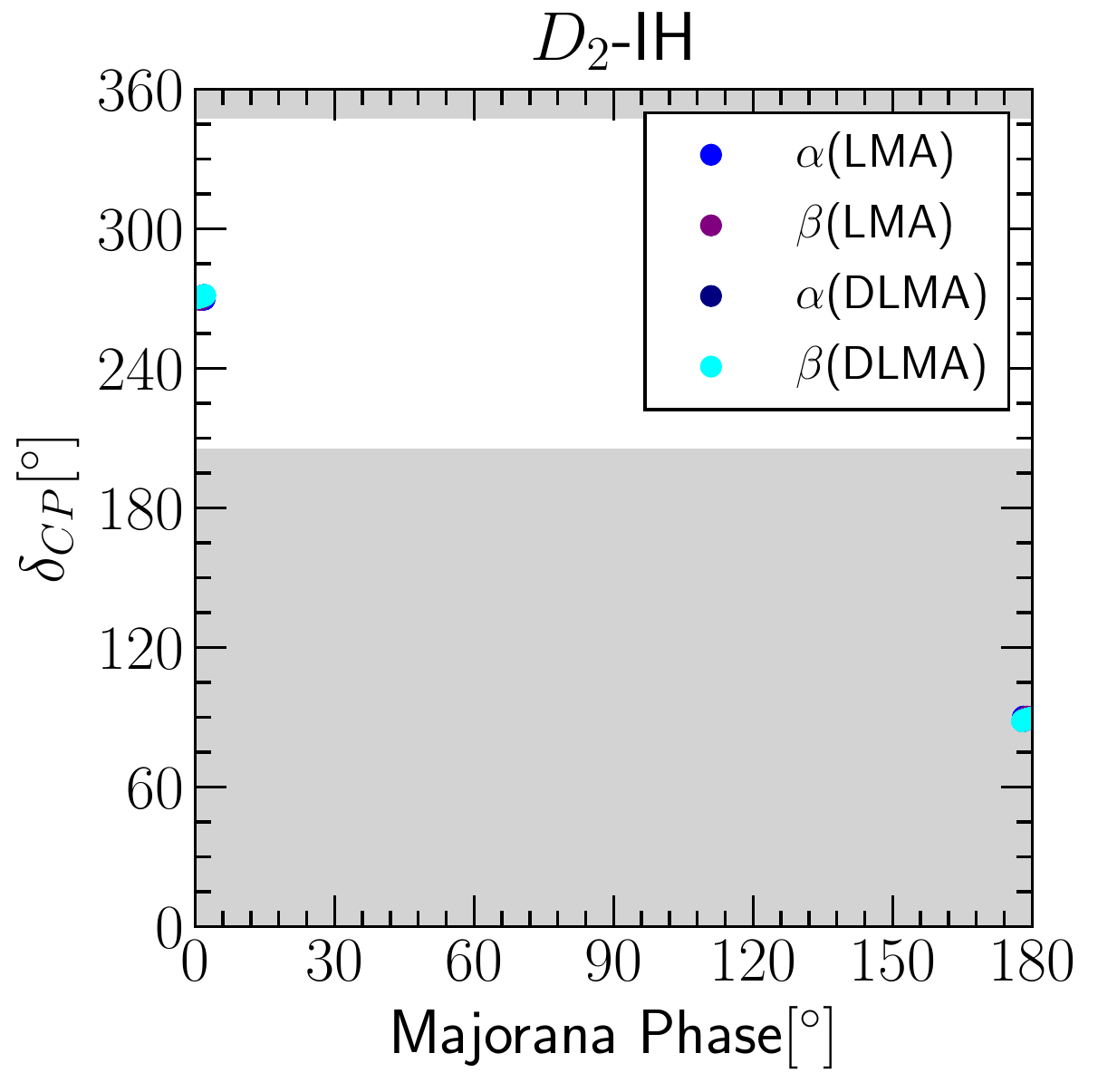}   \\     
\includegraphics[scale=0.37]{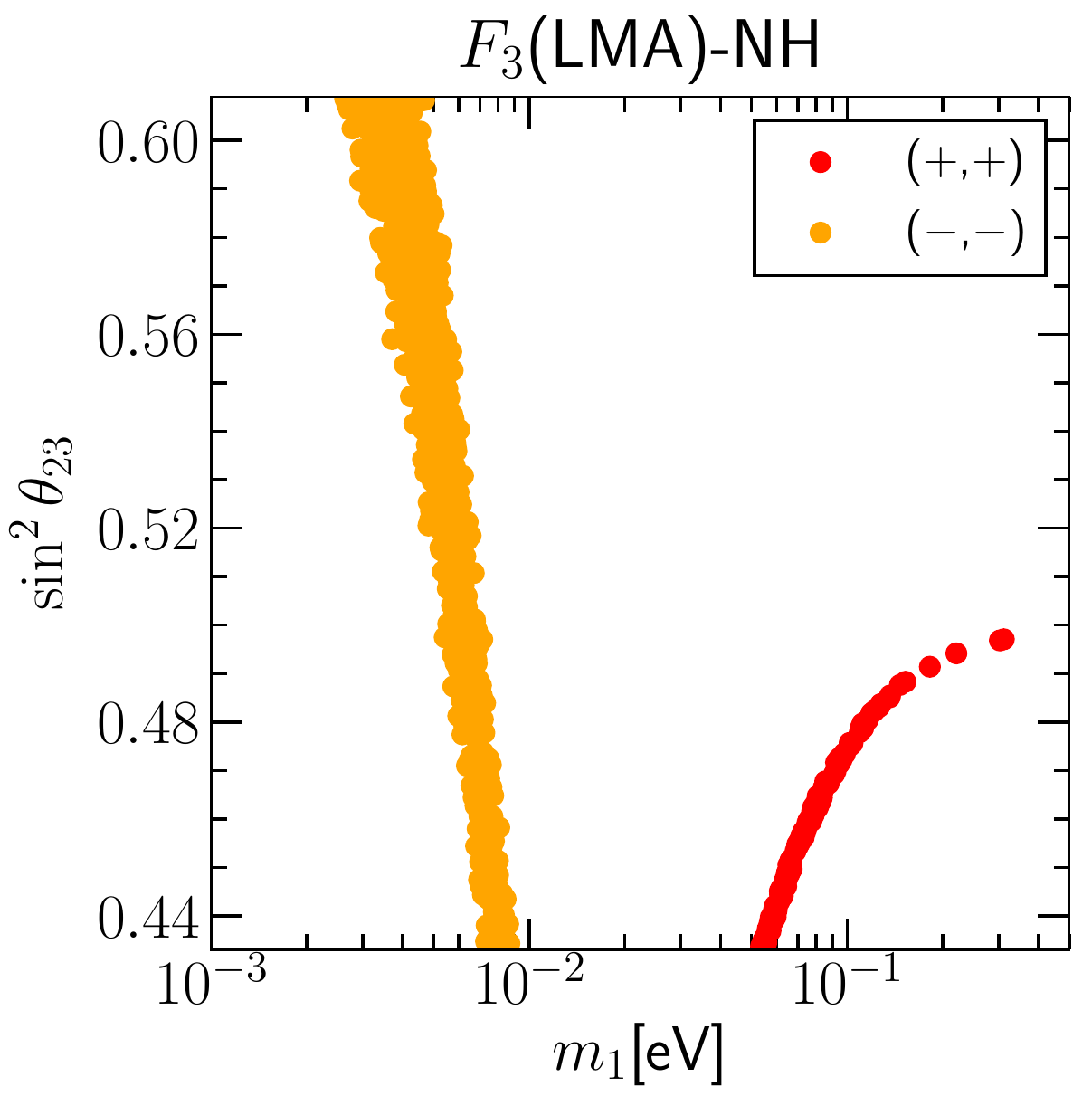}
\includegraphics[scale=0.37]{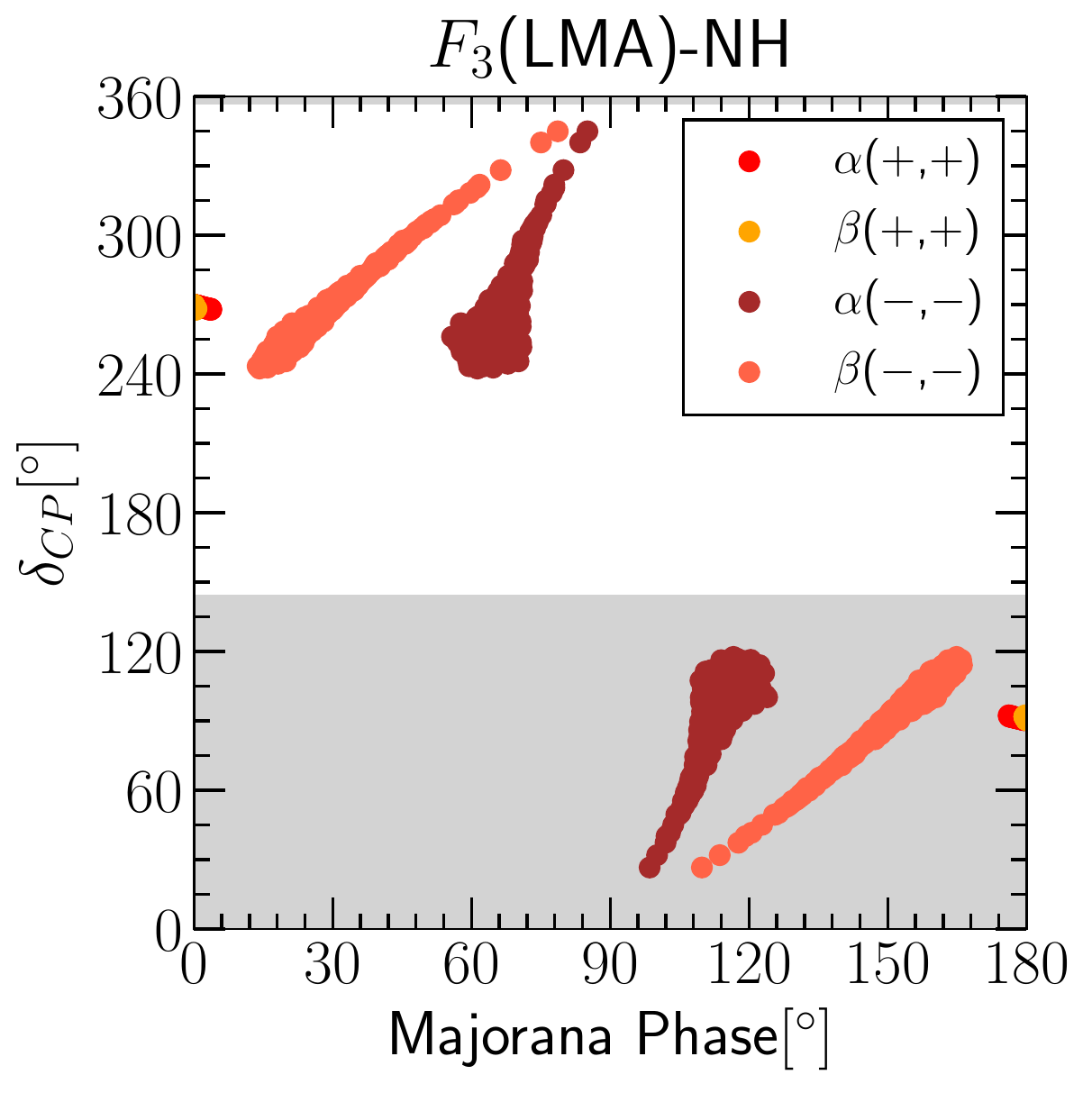}   
\includegraphics[scale=0.37]{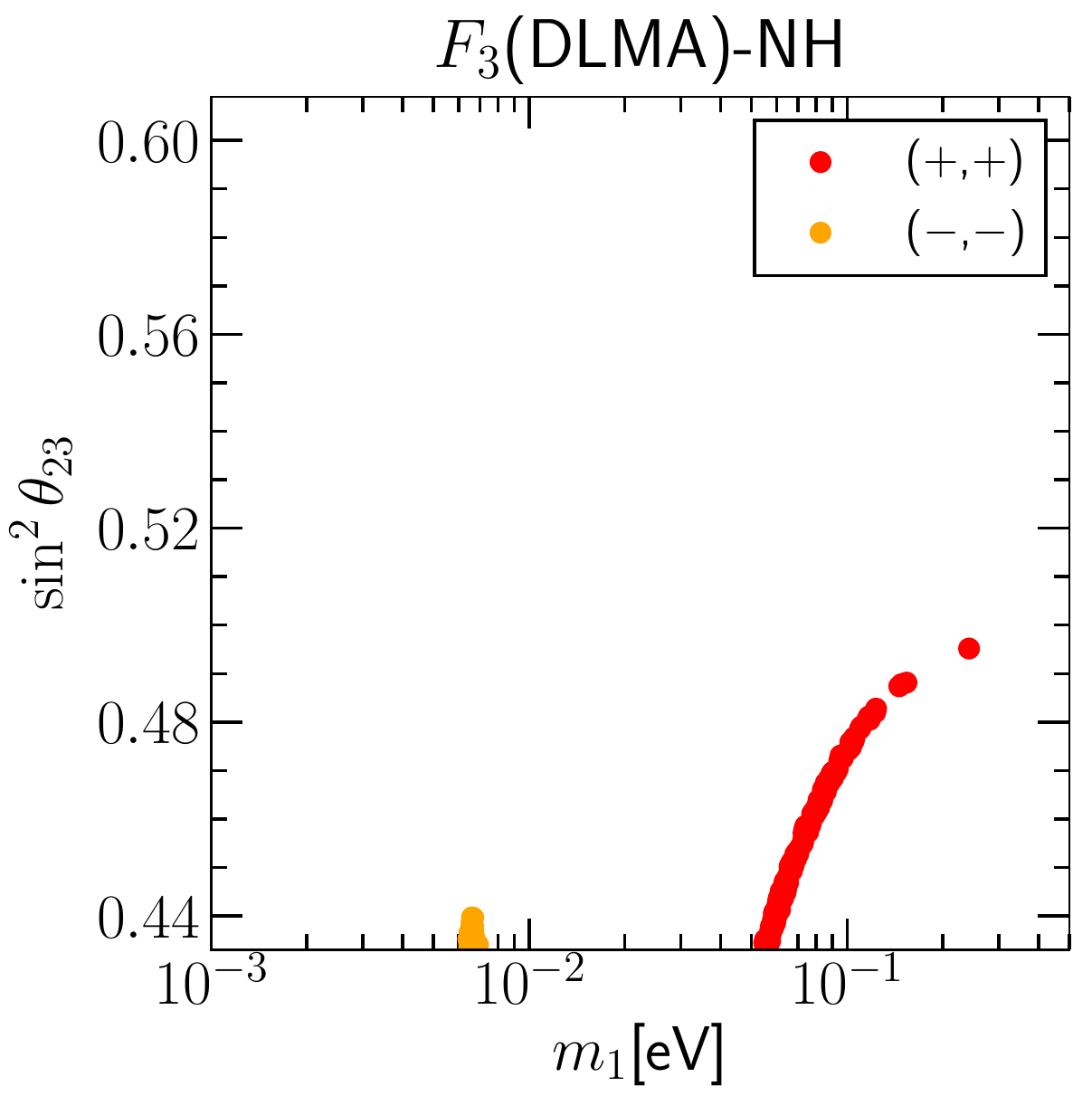} \\
\includegraphics[scale=0.37]{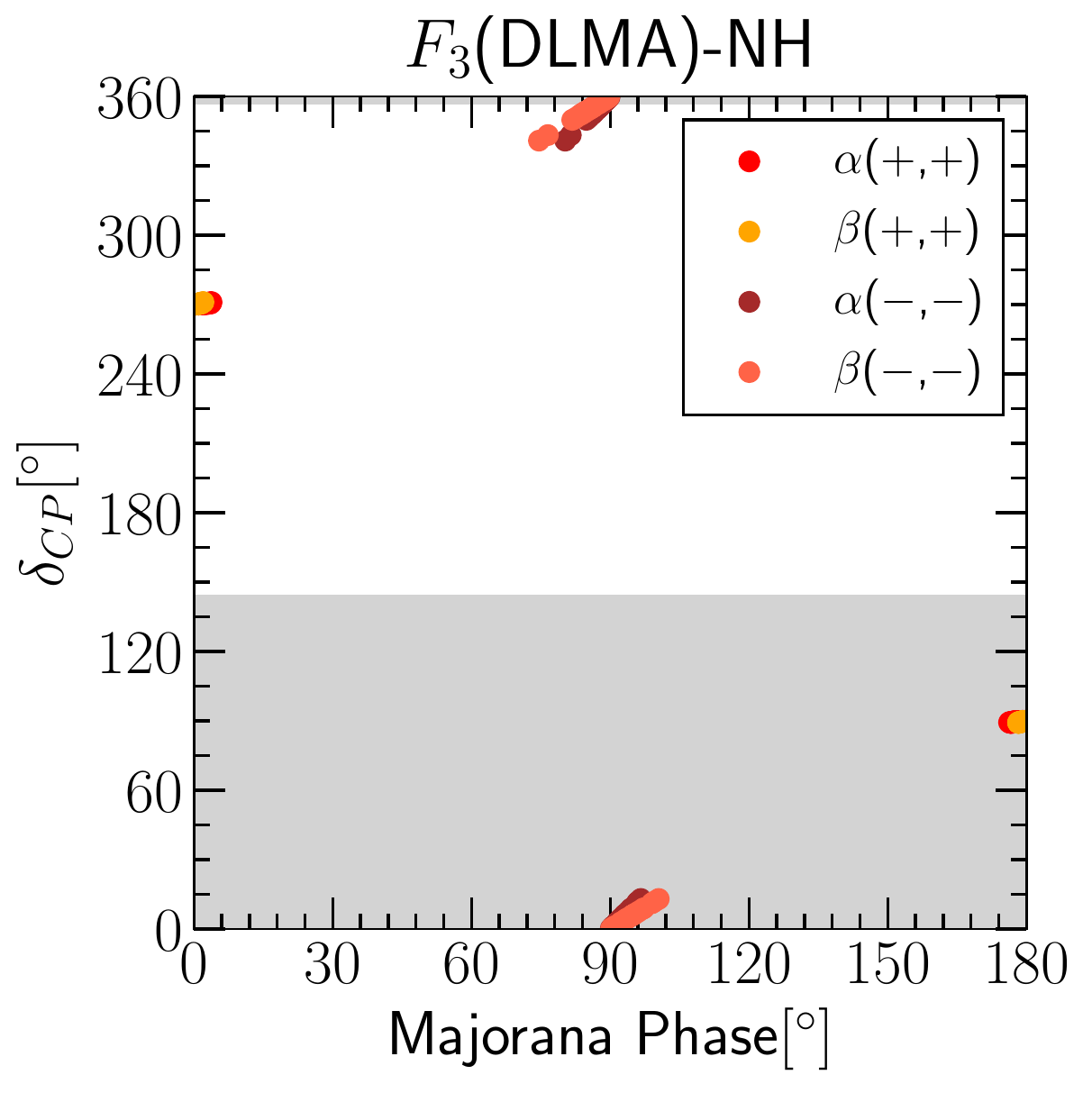}      
\includegraphics[scale=0.37]{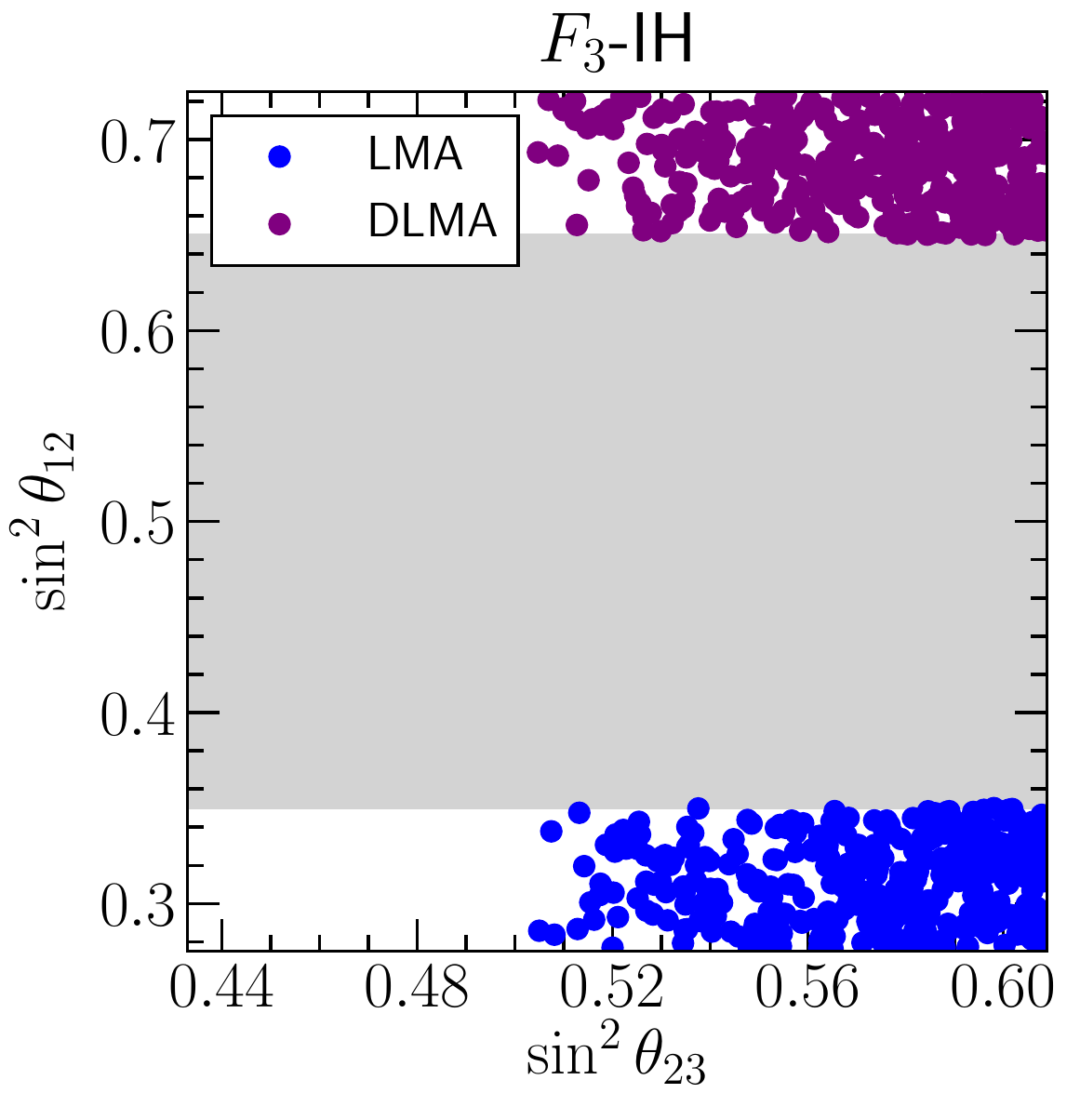}
\includegraphics[scale=0.37]{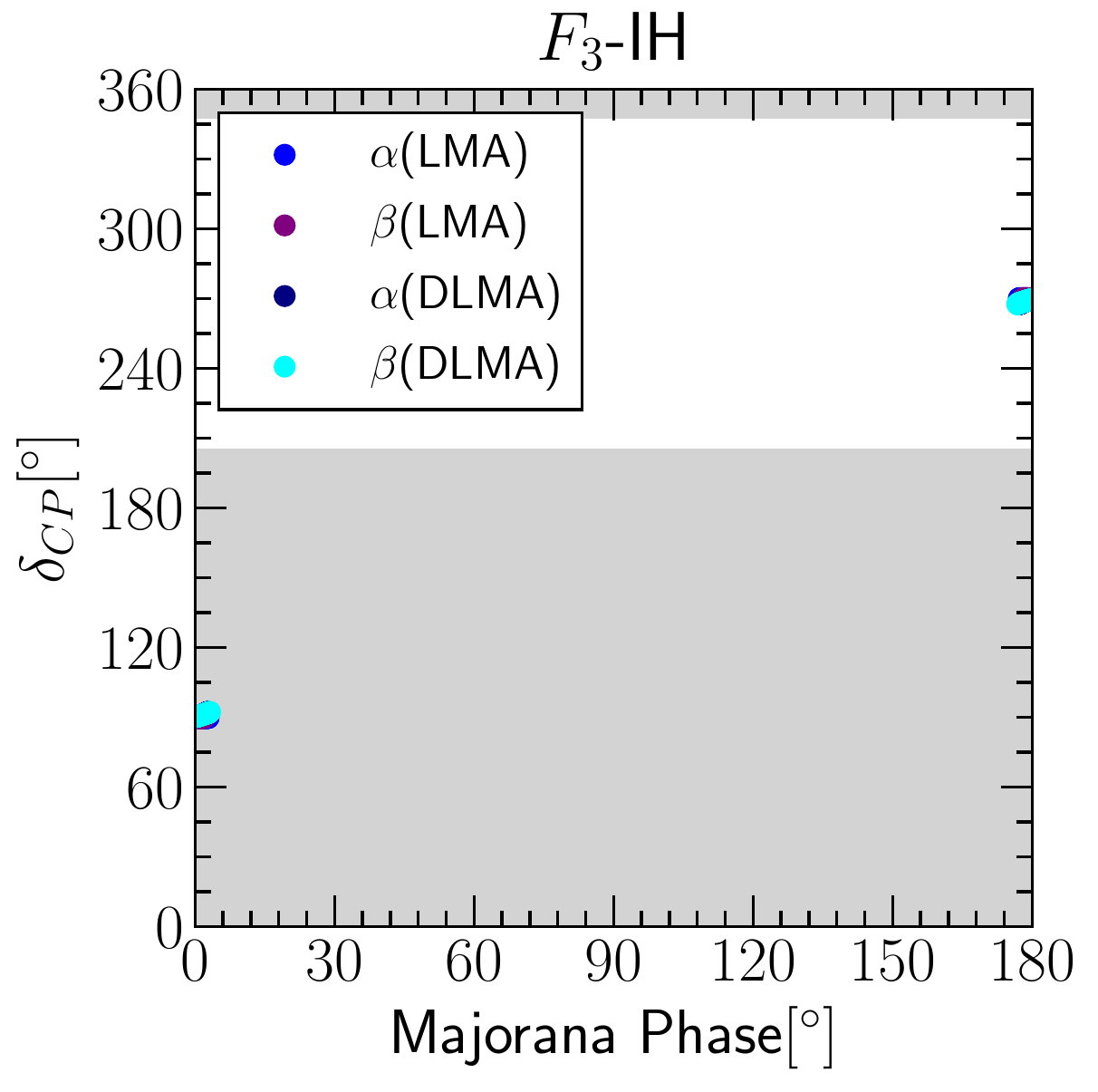}   \\     
\caption{Correlation plots for $D_2$ and $F_3$. The grey shaded areas in these plots are the excluded values of $\delta_{\rm CP}$ and $\sin^2\theta_{12}$ at $3 \sigma$ as obtained by the global fit.}
\label{corr_TC2}
\end{center}
\end{figure*}

\begin{figure*}[h!]
\begin{center}
\includegraphics[scale=0.37]{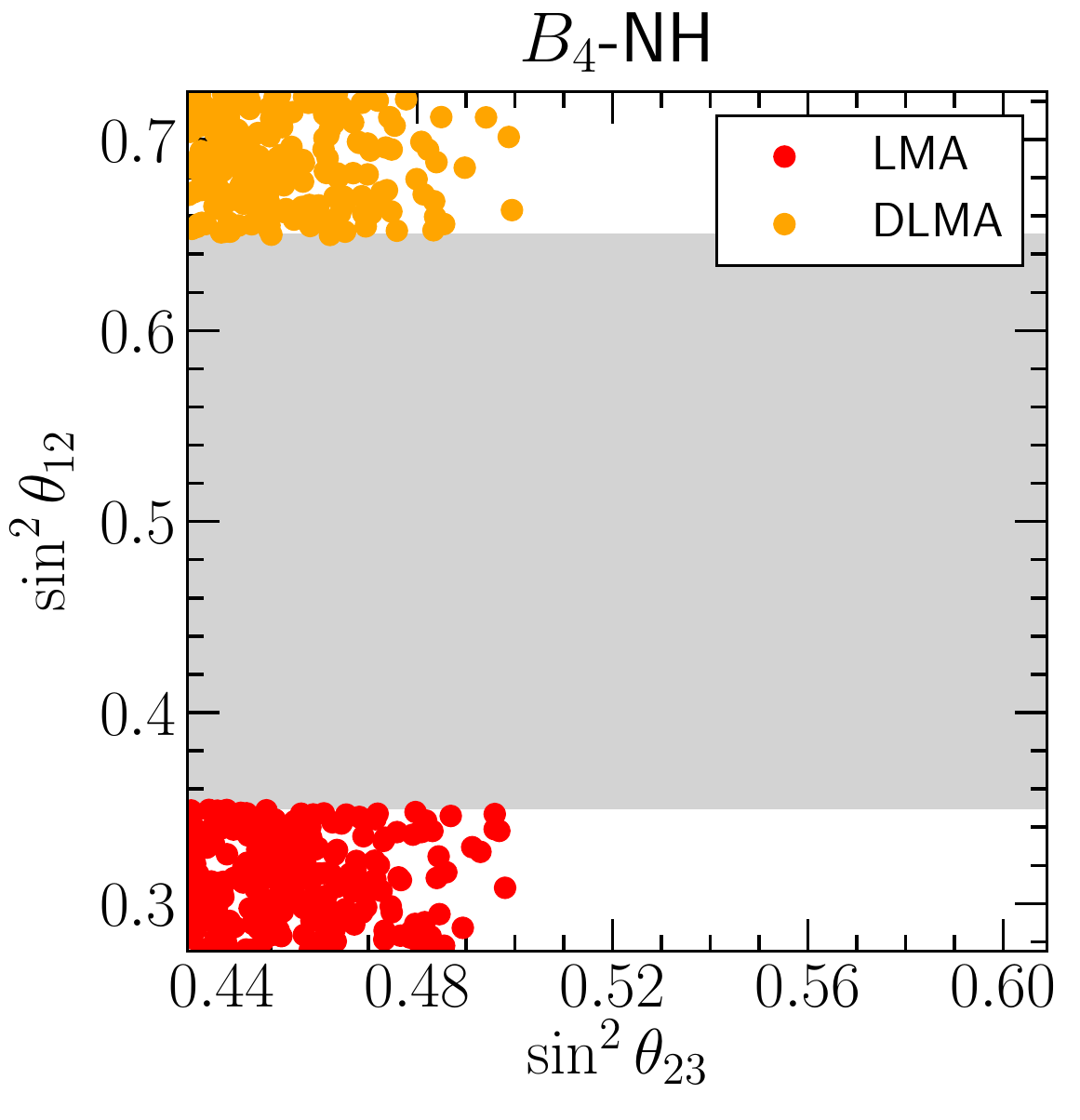}
\includegraphics[scale=0.37]{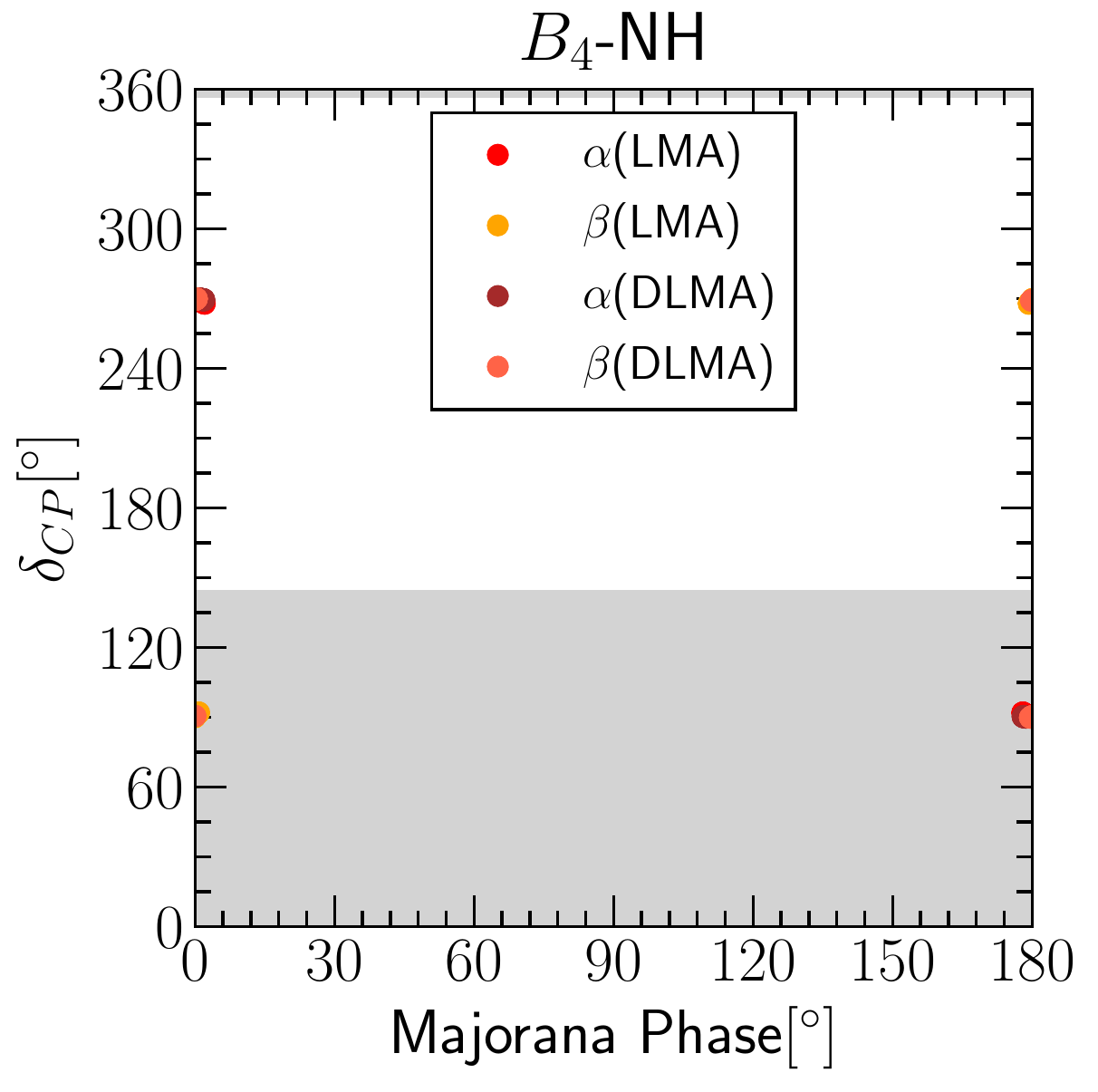}   
\includegraphics[scale=0.37]{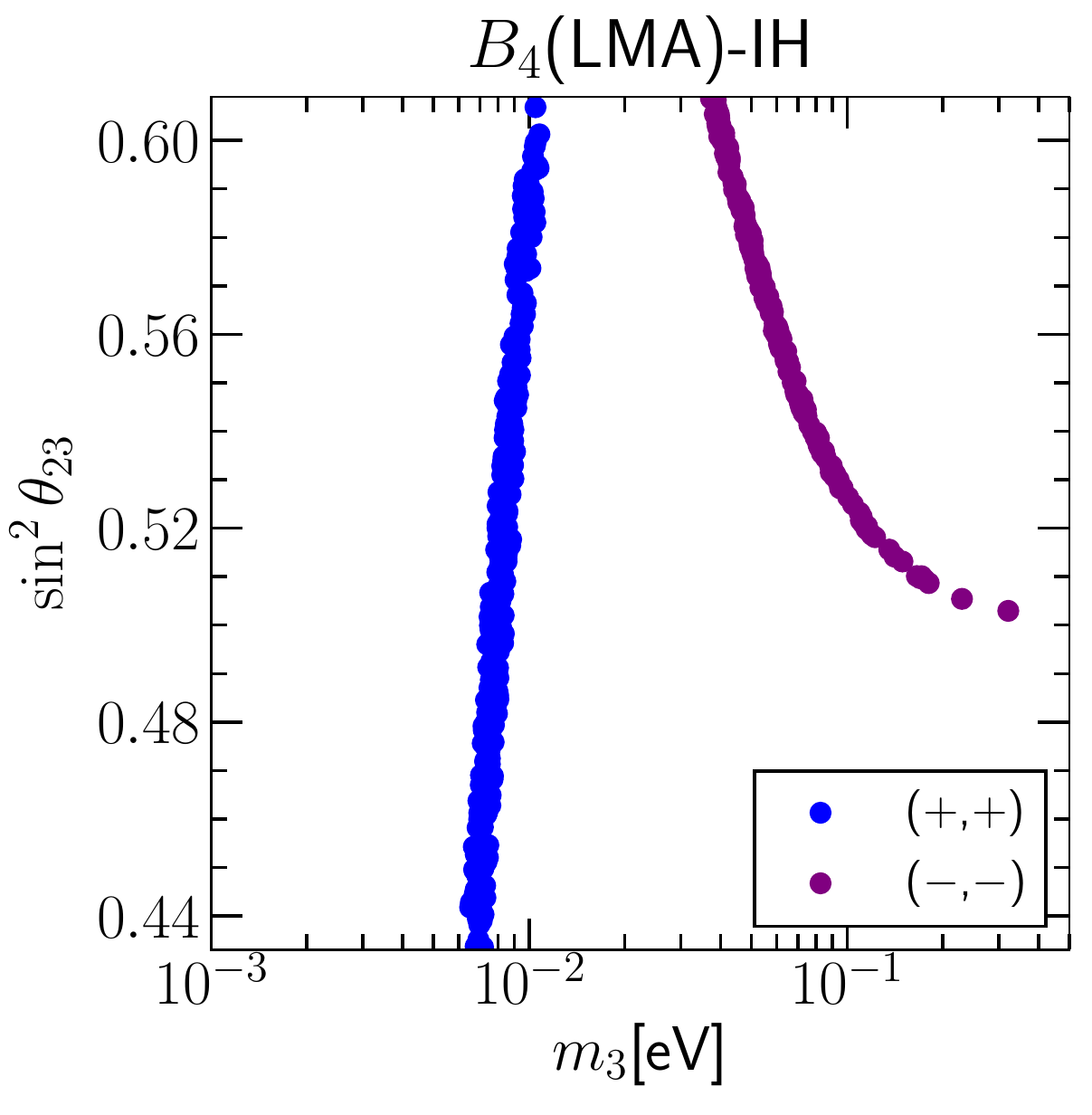} \\
\includegraphics[scale=0.37]{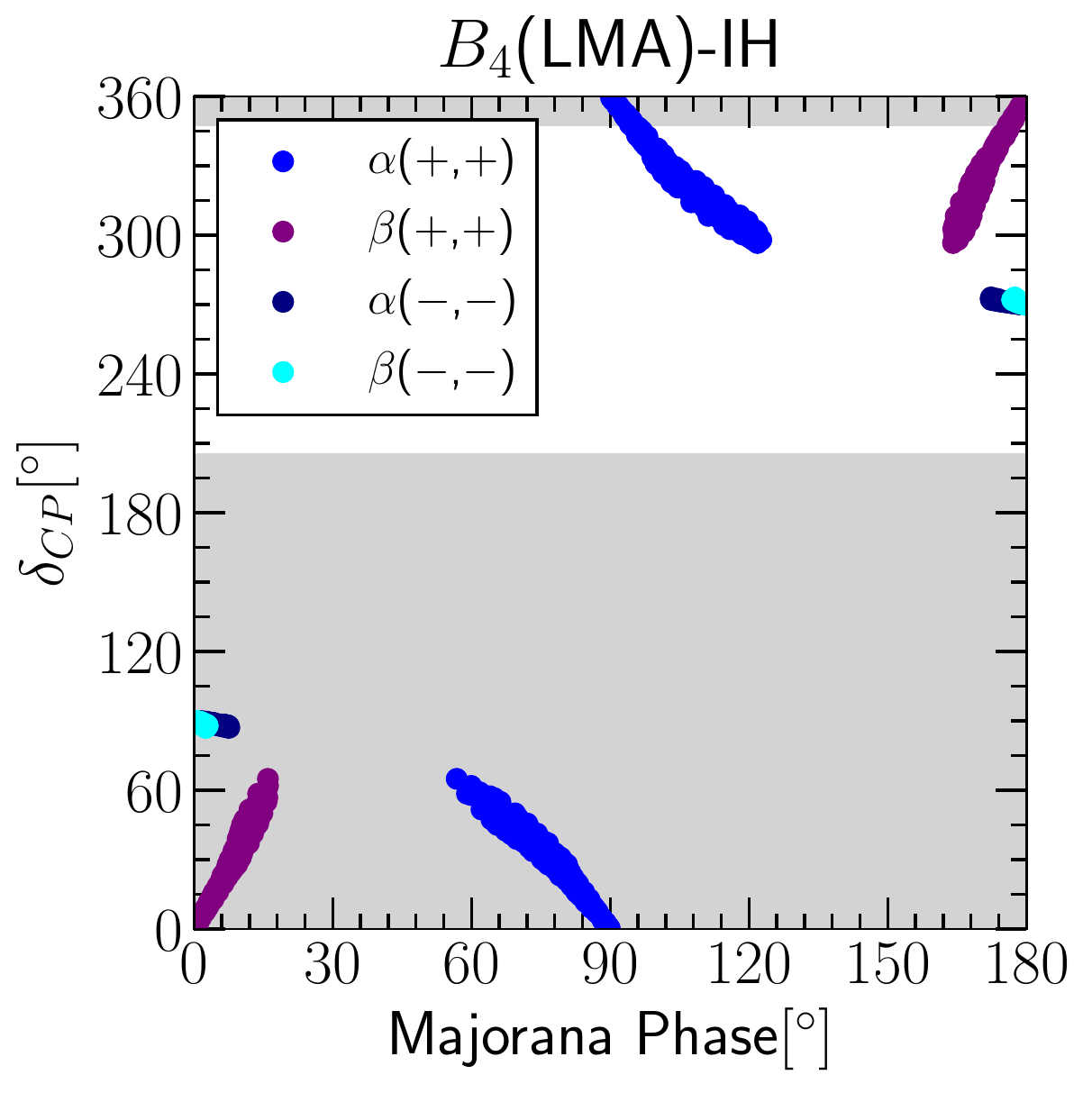}      
\includegraphics[scale=0.37]{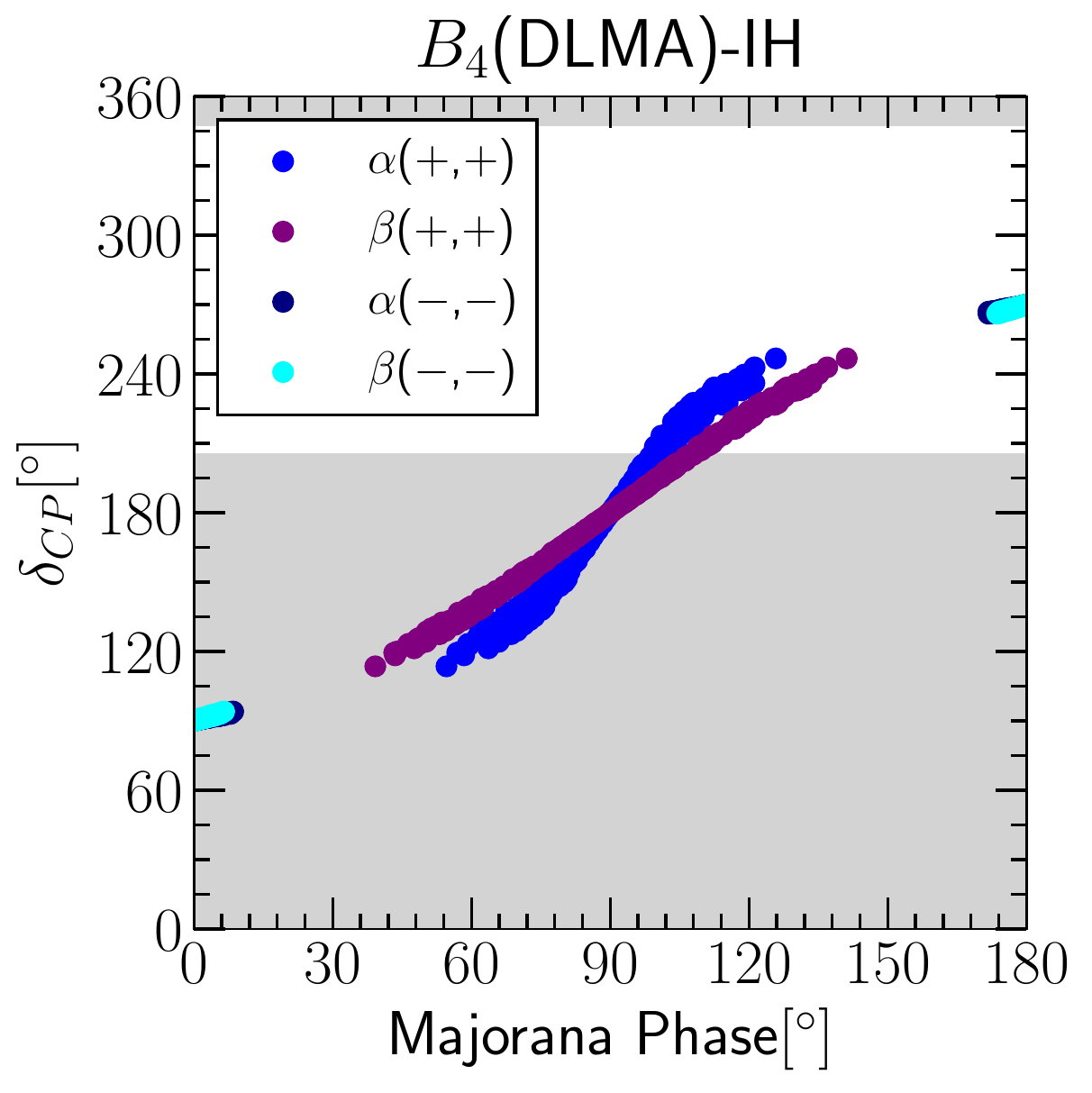}     
\includegraphics[scale=0.37]{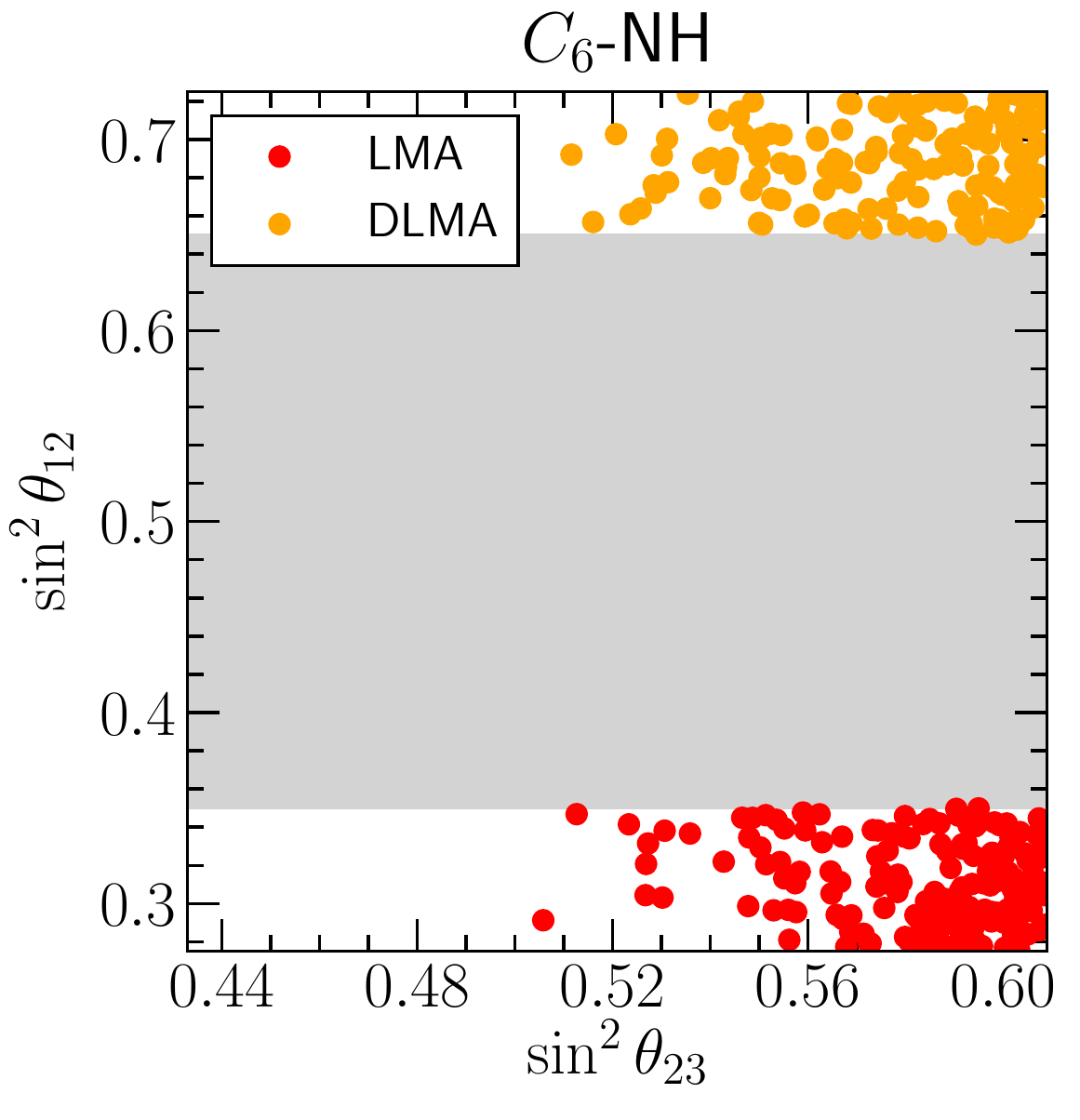} 
\includegraphics[scale=0.37]{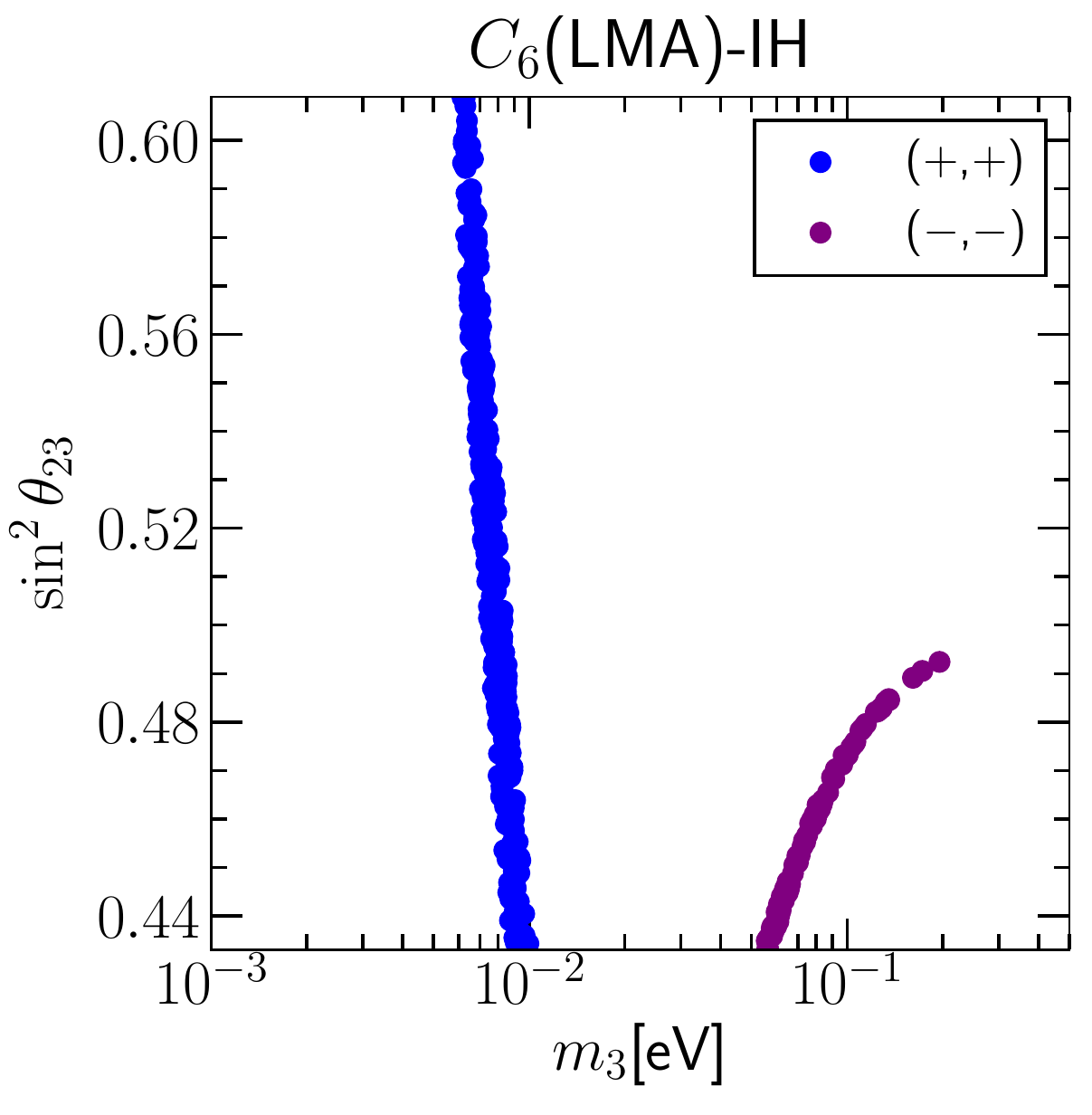} 
\includegraphics[scale=0.37]{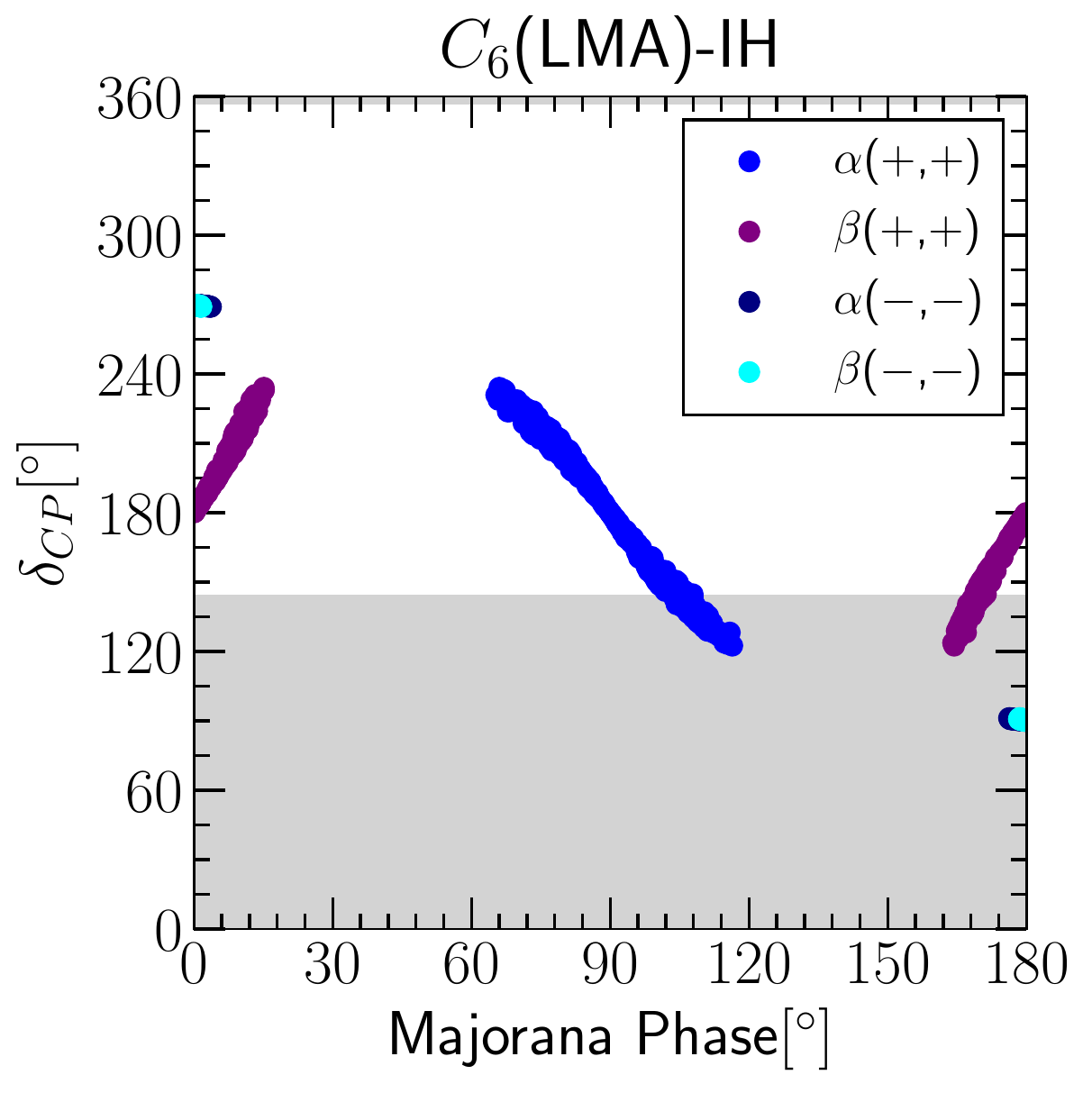}      
\includegraphics[scale=0.37]{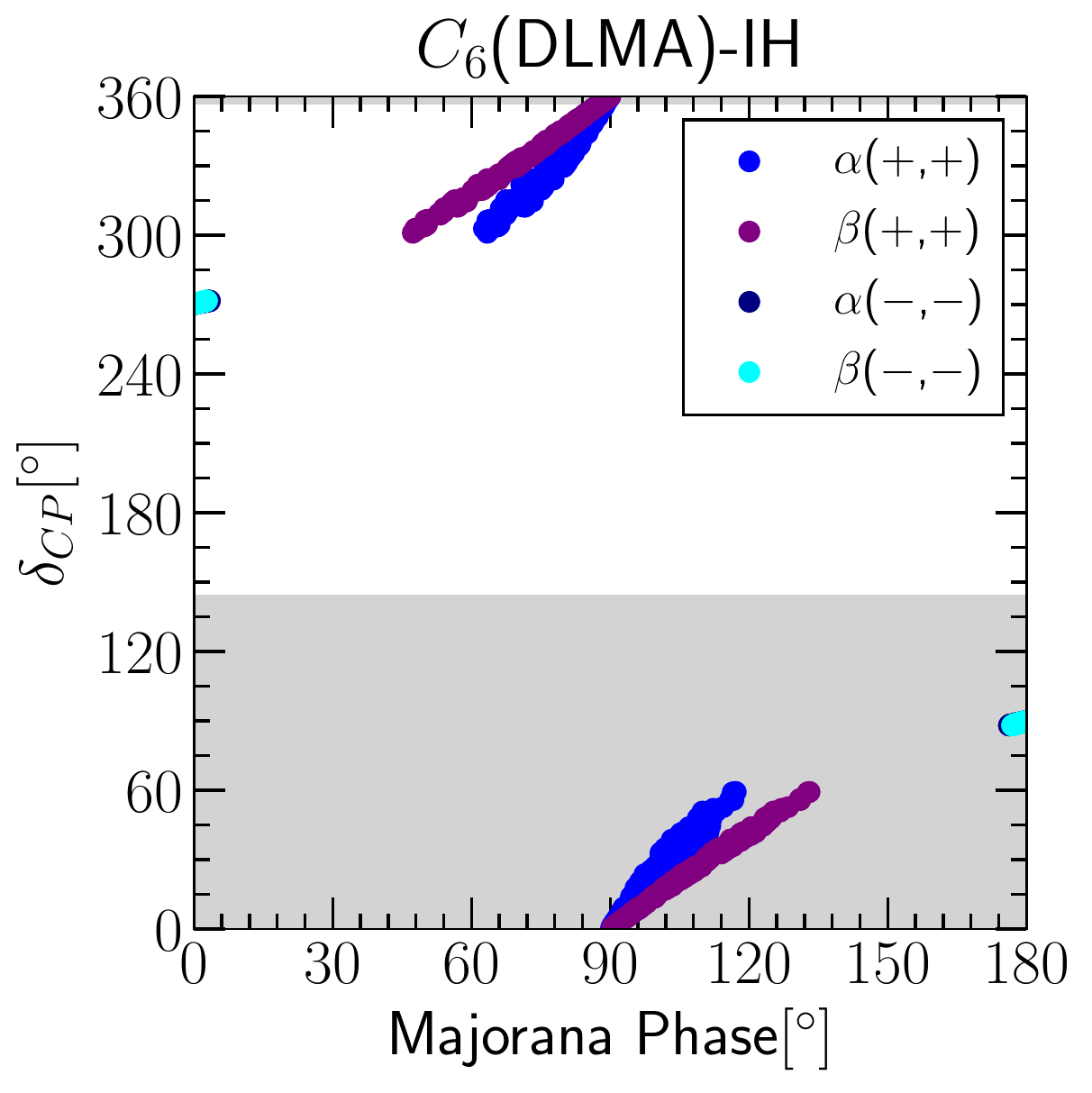}   \\     
\caption{Correlation plots for $B_4$ and $C_6$. The grey shaded area in these plots are the excluded values of $\delta_{\rm CP}$ and $\sin^2\theta_{12}$ at $3 \sigma$ as obtained by the global fit.}
\label{corr_TC3}
\end{center}
\end{figure*}
%\clearpage

\begin{figure*}[t]
\begin{center}
\includegraphics[scale=0.5]{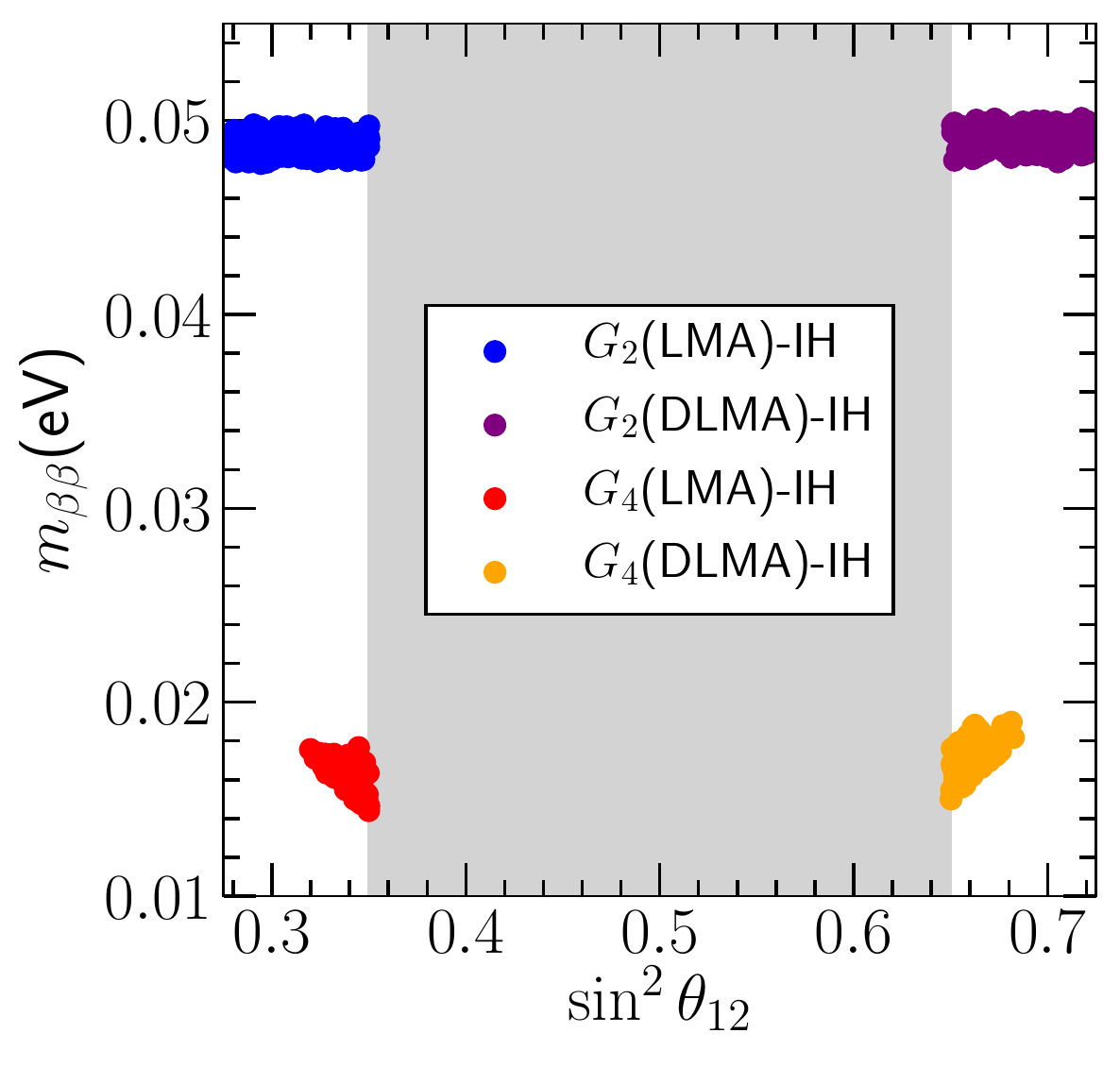}
\includegraphics[scale=0.5]{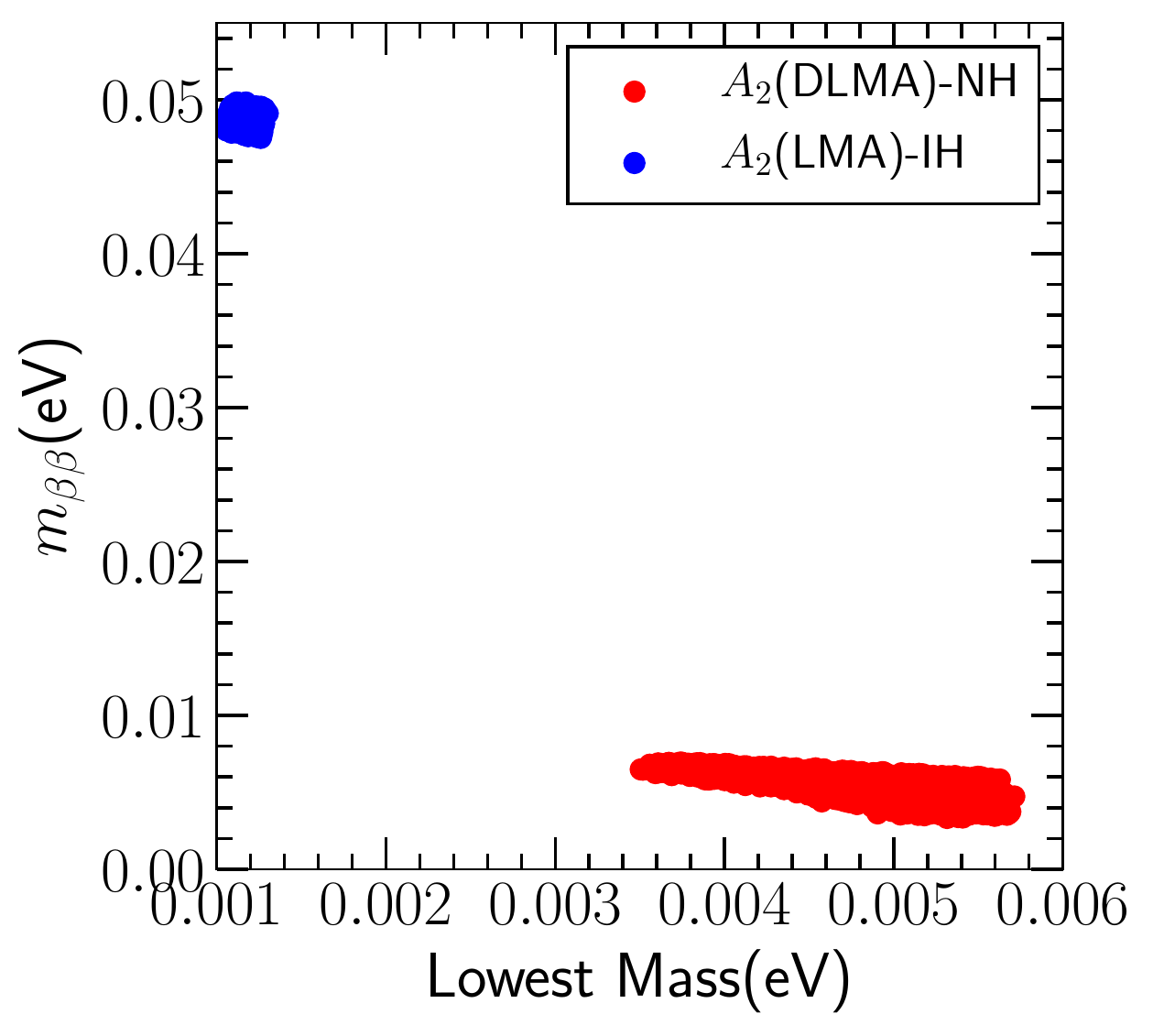}\\ 
\includegraphics[scale=0.5]{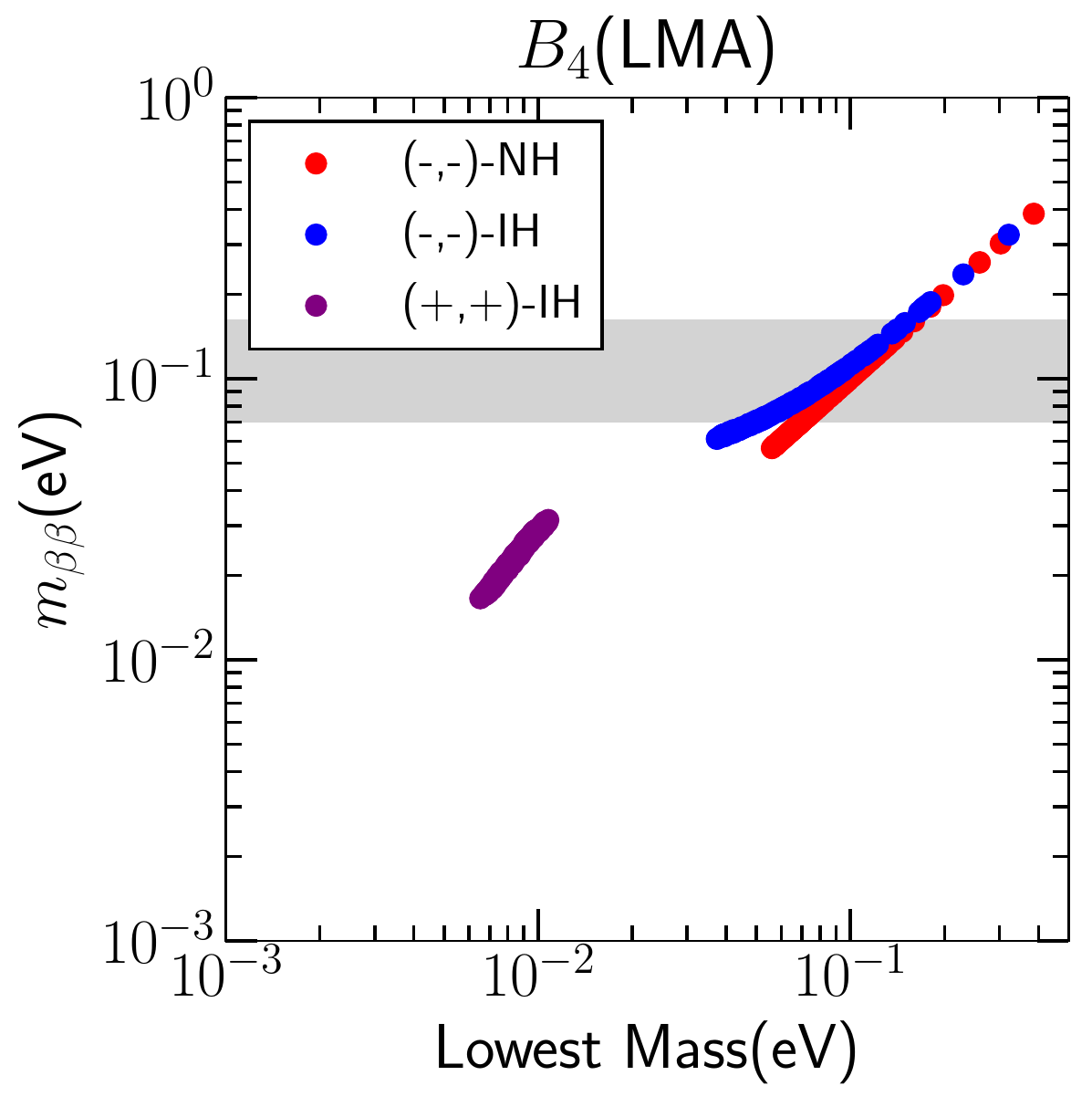} 
\includegraphics[scale=0.5]{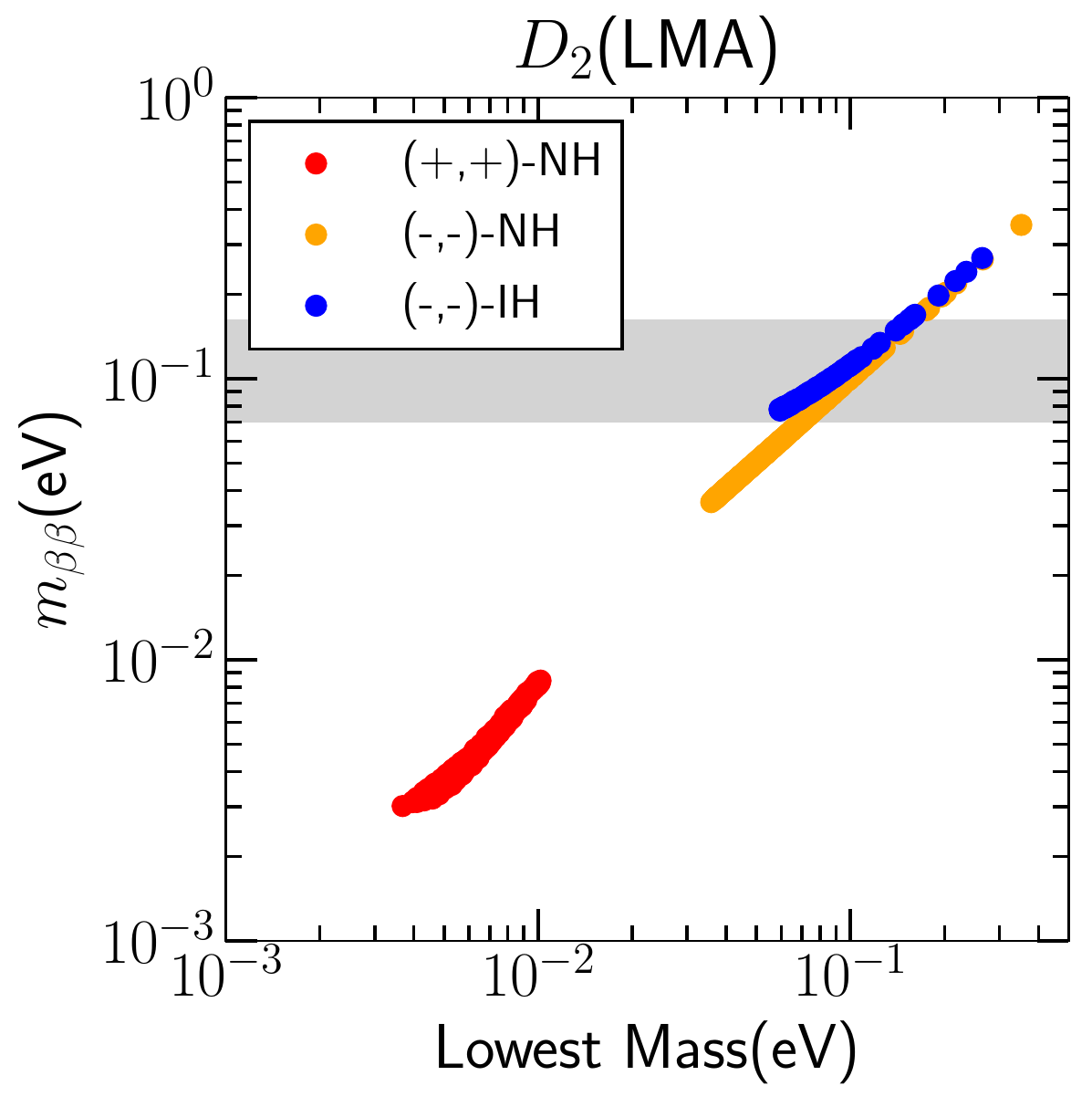} 
\caption{Effective mass prediction versus the neutrino parameters for the allowed texture structures. The grey band indicates the latest KamLAND-Zen bound for effective neutrino mass $|m_{\beta \beta}| \leq (0.061 - 0.165)$  eV \cite{KamLAND-Zen:2016pfg}. For explanation please refer to text.}
\label{ndbd}
\end{center}
\end{figure*}
%\clearpage

\item In Fig.  \ref{corr_TC3} we present the constraints on neutrino observables for the $\mu-\tau$ symmetric class $B_4$ and $C_6$. Here we see that for NH this class remains alive for the $(-,-)$ solution.  
As can be seen from the figure that $D_2$ ($C_6$) class for NH is allowed for both LMA and DLMA provided the true octant for $\theta_{23}$ is the lower (higher) one. Along with this preference, the constraints on the phases for $B_4$ are quite strong allowing a very narrow range of parameter space with maximal $\sin\delta_{\rm CP}$ and almost vanishing $\sin\alpha$ and $\sin\beta$. The correlations for $C_6$ is very similar as that of $B_4$ (not shown).  For IH with the choice of LMA/DLMA we obtain exactly similar correlations between $\sin ^2 \theta_{23}$ and the $m_{\text{lowest}}$ for both $B_4$ and $C_6$ and we present figures only for LMA. For both $B_4$ and $C_6$, the $(+,+)$ solution corresponds to the region $m_1 <  0.02$  eV and the $(-,-)$ solution corresponds to the region $m_1 >  0.02$. For $B_4$ ($C_6$), all the values of $\theta_{23}$ are allowed for $(+,+)$ solution and only higher (lower) octant is allowed for $(-,-)$ solution.
Regarding the constraint on the phases, the $(-,-)$ solution severely restricts the range of the phases with maximal $\sin\delta_{\rm CP}$ and almost vanishing $\sin\alpha~\text{and}~\sin \beta$. This is true for both $B_4$ and $C_6$ and for both LMA and DLMA. 
However for $(+,+)$ solution,  we see that both $B_4$ and $C_6$ with the LMA is allowed for nearly maximal range of $\sin\alpha$  and moderately smaller values for $\sin \beta$ and $\sin\delta_{\rm CP}$. But for the DLMA solution prefers larger values of $\sin \alpha$ and $\sin\beta$ with relatively smaller values of $\sin\delta_{\rm CP}$. 
\end{itemize}

%++++++++++++++++++++++++++++

\section{Predictions for neutrinoless double beta decay}
\label{ndbd}

In the three generation picture the half-life for the $0\nu\beta\beta$ process is given by 
\begin{equation}
\frac{\Gamma_{0\nu\beta\beta}}{\text{ln}2} = G \abs[\Big]{ \frac{ M_{\rm NME}}{m_e}}^2 m_{\beta\beta} ^2
\end{equation}
with $G$ containing the phase space factors, $m_e$ as the electron mass, and $M_{\rm NME}$ as the nuclear matrix element (NME).  One can write the effective neutrino mass ($m_{\beta\beta} \sim M_{ee}$) governing neutrinoless double beta decay process as,
\begin{equation}
m_{\beta\beta} = \abs[\Big]{ \sum_i U_{ei}^2 m_i}
\end{equation}
where $U$ is the unitary lepton mixing matrix and $m_i$ are the mass eigenvalues for the three active neutrinos. Putting the expressions of the mixing matrix elements, the above equation can be recast into the following form
\begin{equation}
m_{\beta\beta}= \abs[\Big]{m_1c^{2}_{12}c^2_{13} + m_2s^2_{12}c^2_{13}e^{2i\alpha}+ m_3s^2_{13} e^{2i\beta}}
\end{equation}
As is evident, the effective neutrino mass depends on all the neutrino oscillation parameters except the atmospheric mixing angle and the CP phase $\delta_{\rm CP}$. Thus, it is expected that any alteration on any of these parameters has significant impact on the effective neutrino mass \cite{PhysRevD.99.095038}. 
Recently Ref. \cite{Borgohain:2020now} has studied the prediction for $m_{\beta\beta}$ for the allowed one zero and two zero textures when the lowest neutrino mass is non vanishing. 
In this section we present the prediction for $m_{\beta \beta}$  for the allowed one zero texture with vanishing lowest neutrino mass and the texture carrying simultaneously the one texture zero with vanishing minor in Fig. \ref{ndbd}. For one-zero texture with a vanishing neutrino mass we obtain allowed solutions only for IH. For this scenario, LMA and DLMA give similar predictions \cite{PhysRevD.99.095038} for $m_{\beta \beta}$. This is reflected in the top left panel of Fig. \ref{ndbd} where we have presented the results for $G_2$ and $G_4$. In addition, the textures which are related by $\mu-\tau$ symmetry also predict similar values of $m_{\beta \beta}$. Therefore, for the one-zero textures with vanishing lowest neutrino mass we have not shown the results for $G_3$ and $G_6$. These values are well below the current bounds on $m_{\beta \beta}$ and can be understood in the following way. For $m_3 = 0$, $m_{\beta \beta}$ is maximum (minimum) when the $\alpha$ is zero (maximal) and this is the case for $G_2$ ($G_4$).

In the upper right panel, we present the predictions for the allowed textures of $A_2$. The prediction for $A_2$(DLMA)-NH is around 0.005 eV while for $A_2$(LMA)-IH, is close to 0.05 eV. The prediction for $A_2$(LMA)-IH is similar with the predictions of $G_2$ because in this case
$m_3$ is small and $\alpha$ is close to zero. $A_2$(DLMA)-IH, also gives similar prediction. As $A_3$ is $\mu-\tau$ symmetric to $A_2$, we have not presented the results for $A_3$.

In the lower panels we present the predictions for $B_4$(LMA) and $D_2$(LMA) for various allowed solutions. For a given hierarchy LMA and DLMA solutions predict similar values of $m_{\beta \beta}$.  In these plots the grey band indicates the latest KamLAND-Zen bound for effective neutrino mass $|m_{\beta\beta}| \leq  (0.061 - 0.165)$ eV \cite{KamLAND-Zen:2016pfg}. For $B_4$ ($D_2$) we notice that the NH (IH) solution is almost ruled out from the current bounds of the $m_{\beta \beta}$. As $F_3$ and $C_6$ are $\mu-\tau$ symmetric to $B_4$ and $D_2$, we have not presented the results for $F_3$ and $C_6$.

Though $m_{\beta\beta}$ depends on $\theta_{12}$, we notice that the predictions for $m_{\beta\beta}$ are same for both LMA and DLMA solutions for all the allowed textures. This can be understood in the following way. For the values $m_1 \simeq m_2 = x$, the expression for $m_{\beta\beta}$ can be re-written as
\begin{eqnarray}
m_{\beta\beta} &=& \abs[\Big]{x c^2_{13} + A}~~~~ {\rm if}~~ \alpha = 0^\circ \label{mbb_alpha_0} \\
                        &=& \abs[\Big]{xc^2_{13} \cos2\theta_{12} + A}~~~~ {\rm if}~~ \alpha = 90^\circ \label{mbb_alpha_90}
\end{eqnarray}
where $A$ is the $\theta_{12}$ independent term containing $m_3$. Note that the Eq. \ref{mbb_alpha_0} is independent of $\theta_{12}$ whereas Eq. \ref{mbb_alpha_90} depends on $\cos2\theta_{12}$ which has same value for both LMA and DLMA solution but with opposite sign. Therefore depending on the values of $A$, the values of $m_{\beta\beta}$ can be almost degenerate in LMA and DLMA solution when $\alpha = 90^\circ$. Now as almost all the allowed textures satisfiy the above conditions, we observe similar prediction for $m_{\beta\beta}$ for both LMA and DLMA solution.

\section{Conclusion}\label{conclusion}

In this work we revisit the texture zeros in low energy neutrino mass matrices in the standard three flavour framework in light of the DLMA solution of the solar mixing angle $\theta_{12}$. In particular we study the cases of (i) one-zero texture and (ii) simultaneous  appearance of one-zero texture and one minor zero. Recently one-zero and two-zero textures in presence of DLMA solution has been performed considering the case when the lowest neutrino mass is non-zero. In this work we re-examined the above two texture classes considering the lowest neutrino mass vanishing. This is motivated by the fact that vanishing lowest neutrino mass corresponds to the most economical extension of the Standard Model.  First of all we find that two zero textures are not allowed if we consider the lowest neutrino mass as zero. For one zero texture, we find that among the six possible scenarios only four are allowed for both LMA and DLMA solution of $\theta_{12}$ in IH. For NH, none of the one zero textures are allowed. For the allowed texture zero cases we have presented the correlation between different mixing parameters and explained our numerical results from the analytical expressions. For the textures with the simultaneous appearance of one-zero texture and vanishing minor we have found that none of the possible cases are allowed when the lowest neutrino mass vanishes. However if the lowest neutrino mass is non-zero then we find that among the 15 possible cases only 6 are allowed. Among the 6 allowed classes, 4 are allowed for both LMA and DLMA solution of $\theta_{12}$. Interestingly there exist two cases for which we find that in the presence of DLMA solution of $\theta_{12}$ they are allowed in NH but they are not allowed with LMA solution of $\theta_{12}$ in NH. In IH, theses two cases are allowed for both LMA and DLMA solution of $\theta_{12}$. Apart from presenting the correlation between different parameters, we have also explained the numerical results of these particular two cases using analytic expressions. Further, we have also presented the predictions for the effective mass $m_{\beta \beta}$ for all the allowed case belonging to the (i) one zero textures considering lowest neutrino mass to be zero and (ii) simultaneous appearance of texture zero and vanishing minor when the lowest neutrino mass is non-zero. We identified two classes belonging to the case of simultaneous appearance of texture zero and vanishing minor, which can be excluded from the current bounds on the $m_{\beta \beta}$ from the neutrinoless double beta decay experiments. We expect that future neutrino experiments will provide us more accurate determination on the true octant of the atmospheric mixing angle, CP-violating phases, mass hierarchy and will finally guide us to choose the appropriate form for low energy neutrino mass matrix.

%\clearpage
%=======================
\bibliographystyle{JHEP}
\bibliography{texture_dlma}
%=======================

%============
\end{document}